\pdfoutput=1
\documentclass[aps,prb,twocolumn,showpacs]{revtex4}

\usepackage{t1enc}
\usepackage{amsmath}
\usepackage{graphicx}
\usepackage{epstopdf}
\usepackage{color}

\begin{document}
\title{Collective diffusion of  dense  adsorbate at surfaces  of  various geometry }

\author{Marcin Mi{\'n}kowski, Magdalena A. Za{\l}uska--Kotur}
\email{minkowski@ifpan.edu.pl,zalum@ifpan.edu.pl}
\address{Institute of Physics, Polish Academy of Sciences, Al.~Lotnik{\'o}w 32/46, 02--668 Warsaw, Poland}

\begin{abstract}
Convenient variational formula for collective diffusion of many particles adsorbed at lattices of arbitrary geometry is formulated. The approach allows to find the expressions for the diffusion coefficient for any value of the system's coverage. It is assumed that particles interact via on-site repulsion excluding double site occupancy. It is shown that  the method can be applied to various systems of different geometry. Examples of real systems such as GaAs with specific energetic landscapes are also presented. Diffusion of Ga adatoms on GaAs(001) surface reconstructed in two different symmetries is studied. It is shown how increasing Ga coverage changes the character of diffusion from isotropic two-dimensional into highly anisotropic, almost one-dimensional. It is shown how important the role of the inter-particle correlations is, which influence the value of the collective diffusion coefficient at higher coverages.
\end{abstract}
\maketitle
\section{Introduction}
Collective diffusion is a complex many-body process that plays an important role in many physical applications. During epitaxial crystal growth interplay between adsorption, desorption and surface diffusion is responsible for the character and the quality of the grown layers. Understanding diffusive processes is therefore important to gain a precise control over the crystal growth. Currently of particular interest is synthesis of atom-thick layers such as graphene \cite{Li1,Bonaccorso}, semiconductors like MoS$_{2}$ \cite{Lee} or hexagonal-BN \cite{Ismach}, or topological insulators such as Bi$_{2}$Se$_{3}$ \cite{Yan,Li2} or Bi$_{2}$Te$_{3}$ \cite{Li2}.

The value of the diffusion coefficient (or, in general, the diffusion tensor), which can be either measured in experiments or calculated using theoretical methods, strongly depends on the system's energy landscape, which describes the depth of all the adsorption sites and the jump rates between them. Specifically, both the rate of diffusion and its anisotropy depend on the details of the energetic landscape. Those parameters play an important role during growth of both bulk crystals and various nanostructures at their surfaces.

Even though crystals usually crystallize in highly symmetric structures, the symmetry of their surfaces is often different from that of the bulk. One of the reasons for it is surface reconstruction, which can lead to a change of the height of the energy barriers \cite{Matrane} and the geometry of the surface \cite{Roehl1,Meier}. Other reasons can be presence of steps at the surface \cite{Cai,Yildirim} or defects \cite{Tosoni}.

Theoretical calculations of diffusion coefficients require the knowledge of the system's energetic landscape, which is usually determined from ab-initio methods or from experiments. Among common simulation approaches to diffusion  problems are Monte Carlo and molecular dynamics, while analytical methods are based for example on Fokker-Planck, Kramers or master equation.  \cite{Ala-Nissila,Zwerger,Reed,Gomer,Gartner,Danani,Mussawisade,Tarasenko,Chvoj,Medved,Payne,Gosalvez1,Gosalvez2,Gortel,Zaluska-Kotur1,Zaluska-Kotur2,Krzyzewski,Minkowski1,Minkowski2,Minkowski3,Minkowski4,Minkowski5}. Many studies have been devoted to the problems of diffusion over concrete crystal surfaces of complex energetic landscape \cite{Gosalvez1,Gosalvez2,Minkowski2,Minkowski4,Minkowski5}. In such case description of single-particle diffusion already is a difficult task and the presence of many particles increases only the degree of complexity.

Our approach to calculating diffusion coefficients is a variational one and is based on analysis of master equation. It was first proposed in Ref. \onlinecite{Gortel} and has been subsequently developed to deal with various cases of surface diffusion \cite{Minkowski1,Zaluska-Kotur1,Zaluska-Kotur2,Krzyzewski,Minkowski2,Minkowski3,Minkowski4,Minkowski5}. It has been shown that the approach is effective in analyzing diffusion in real systems, such as Ga adatom at GaAs(001) surface \cite{Minkowski2} or Cu monomer and dimer on Cu(111) and Ag(111) surfaces \cite{Minkowski4}. Recently using the variational approach we have derived a general formula for the collective diffusion coefficient for interacting particles in an arbitrary one-dimensional potential \cite{Minkowski5}.

For a small concentration of particles in the system their diffusive paths can be treated as independent and the collective diffusion coefficient can be considered as being equivalent to that of single-particle diffusion. In Refs \onlinecite{Gosalvez1} and \onlinecite{Gosalvez2} analytical formulas for single-particle diffusion coefficients for various lattices of square, rectangular and hexagonal symmetry have been derived. It is known that collective diffusion in an energetic landscape consisting of adsorption sites of equal depth does not depend on the coverage if only site blocking interactions are considered \cite{Kutner}. Therefore, in such cases collective diffusion is equivalent to that of a single particle in the same potential. In Ref. \onlinecite{Minkowski1} we  presented  diffusion coefficients for a few lattices of square and hexagonal symmetry with identical adsorption energies calculated using  variational method. 
However, in the case of energetic landscapes with adsorption sites of different depths the collective diffusion coefficient may significantly depend on the system's coverage even if no other interactions than site blocking are present. The degree to which the diffusion is influenced by the inter-particle correlations depends on the difference between the energies of the adsorption sites, the heights of the diffusive barriers and the number of the nearest neighbors, that is the geometry of the system.

In this article we show how to apply the variational approach in calculation of diffusion coefficient of 2D many-particle systems. We calculate collective diffusion coefficients for several systems. The approach that we show here is general and can be applied in any other case, regardless of the symmetry of the network. Among various examples we study also lattices considered in Refs \onlinecite{Gosalvez1} and \onlinecite{Gosalvez2}, for which single-particle diffusion coefficients were calculated. We show density dependence of the diffusion coefficient for all systems studied by taking into account the role of the inter-particle correlations. In the low density limit our results reproduce formulas obtained for single-particle diffusion.

Next  we  use the presented  method to calculate  density  dependence of  the diffusion  on GaAs surface,  studied before in Ref. \onlinecite{Minkowski2}. We  show  that the  diffusion  anisotropy  at  this  surface  can  be  induced  by the  presence   of  other  particles. Density dependent particle kinetics on  lattices of  two  different  symmetries resulting    from different surface  reconstructions  are  compared. 

\section{Theory}
\subsection{Definitions}
We consider a gas of particles adsorbed at a lattice of arbitrary geometry. The lattice is periodic and consists of elementary cells, each of which contains several adsorption sites. In the system we assume periodic boundary conditions. From the physical point of view adsorption sites are energetic minima of the potential landscape that comes from the underlying surface. We assume that at any time a given adsorption site can be either free or occupied by exactly one particle, that is multiple occupancy is forbidden. If there are no interactions between particles, then the occupancy of a given site is independent of the occupancies of the other sites. In the equilibrium the probability that the site with the adsorption energy $E_{A}$ is occupied is
\begin{equation}
\theta_{A}=\frac{ze^{-\beta E_{A}}}{1+ze^{-\beta E_{A}}}\label{eq_prob}
\end{equation}
in a grand canonical ensemble, where $z=e^{\beta\mu}$, $\mu$ is the chemical potential and $\beta=1/k_{B}T$ is the Boltzmann factor. The mean coverage of the system is defined as
\begin{equation}
\theta=\frac{\sum_{i=1}^{n}\theta_{i}}{n},
\end{equation}
where the sum is performed over all the adsorption sites in an elementary cell and $n$ is the number of those sites.

Most of the time the adsorbed particles reside at their sites but from time to time they perform thermally activated jumps to neighboring sites. The jumps occur over the energetic barriers, which are the saddle points of the potential landscape between the two involved adsorption sites. The jump rate from the site A to the site B is
\begin{equation}
W_{AB}=\nu\exp(-\beta(E_{AB}-E_{A}))\label{jump_rate}
\end{equation}
where $E_{A}$ is the adsorption energy of the site A and $E_{AB}$ is the energy of the saddle point between the sites A and B. The prefactor $\nu$ is the attempt frequency, which is the measure of how many attempts per unit time the adsorbed particle performs to jump out of its site. Its value is determined by the vibrations of the atom lattice, that is phonons. Usually we will assume that it is equal to the typical phonon frequency in condensed matter, which is $\nu=10^{13}/s$. At thermal equilibrium detailed balance condition is fulfilled, which means that the probability of every microscopic process is equal to the probability of the reverse process. For our system that condition can be written as
\begin{equation}
W_{AB}\theta_{A}(1-\theta_{B})=W_{BA}\theta_{B}(1-\theta_{A}).\label{det_bal}
\end{equation}
It is easy to see that (\ref{eq_prob}) and (\ref{jump_rate}) satisfy the above condition.

Diffusion is the process of random jumps of the particles by which the system approaches the thermal equilibrium. We will consider diffusion that takes place close to the equilibrium, that is when there are only small deviations from that state. Such situation occurs for example during epitaxial crystal growth.
\subsection{Variational approach}
In order to calculate collective diffusion coefficients for the lattices considered in the next chapter we use the variational approach that was first proposed in Ref. \onlinecite{Gortel}. For one-dimensional collective diffusion it was possible to derive by means of that approach a general formula for the diffusion coefficient expressed in terms of jump rates, interaction constants, temperature and chemical potential \cite{Minkowski5}. Diffusion in one dimension can be considered as an extreme case of anisotropic surface diffusion. There is always only one channel of diffusion in one dimension since a diffusing particle must overcome all barriers ordered one after another to move to the next elementary cell. However, due to inter-particle correlations the dependence of the diffusion coefficient on the barriers is still not trivial and we showed that its behavior can depend not only on the height of those barriers but also on their order in an elementary cell.

Diffusion in two dimensions is a much more complicated process since there are a lot of different paths along which a particle can move. Because of that the diffusion depends not only on the values of the jump rates but also on the geometry of the lattice. We shall derive a variational formula for a non-interacting case, which depends on a set of variational parameters and can be explicitly written out for a system of arbitrary geometry. Then it can be used for finding the values of the variational parameters and therefore determining the diffusion coefficient.

The starting point in our variational approach is the master equation written for the lattice gas system described in the previous section
\begin{equation}
\frac{d}{dt}P(\{c\},t)=\sum_{\{c'\}}[W(\{c\},\{c'\})P(\{c'\},t)-W(\{c'\},\{c\})P(\{c\},t)],\label{master_eq}
\end{equation}
which describes the evolution of the probability of finding the system in the microstate $\{c\}$. A microstate is defined as
\begin{equation}
\{c\}=[\vec{X};\vec{l}_0;\vec{m}_1,\vec{m}_2,\ldots,\vec{m}_{N-1}]=[\vec{X};\{\vec{m}\}],
\end{equation}
which specifies which of the adsorption sites at the lattice are occupied. It distinguishes one of the particles as the reference particle at the position $\vec{X}+\vec{l}_0$, where $\vec{X}$ is the position of the origin of its elementary cell and $\vec{l}_0$ is its position within that cell. The set of the remaining vectors $\{\vec{m}_{1},\vec{m}_{2},\ldots,\vec{m}_{N-1}\}$ gives the positions of all the other particles with respect to the reference particle. Since the system has translational symmetry we can define a configuration as
\begin{equation}
\{\vec{m}\}=[\vec{l}_0;\vec{m}_1,\vec{m}_2,\ldots,\vec{m}_{N-1}],
\end{equation}
which specifies only the position of the reference particle with respect to the origin of any elementary cell and the positions of the remaining particles with respect to the reference particle. All the microstates in which the positions of the particles differ only by a linear combination of the lattice vectors (that is by $\vec{X}$) correspond to the same configuration.

We can use the definition of configuration to replace $W(\{c\},\{c'\})$ in Eq. (\ref{master_eq}) by $W_{\{\vec{m}\},\{\vec{m}'\}}$. At given time we allow only for one jump of a particle between neighboring sites. For pairs of $\{\vec{m}\}$ and $\{\vec{m}'\}$ which fulfill that condition $W_{\{\vec{m}\},\{\vec{m}'\}}$ are equal to corresponding jump rates (\ref{jump_rate}) and in the equilibrium $P(\{c\},t)=P_{\{\vec{m}\}}(\vec{X},t)$ become $P^{eq}_{\{\vec{m}\}}$, which for non-interacting particles is equal to products of $\theta_{i}$ and $1-\theta_{i}$, whose values are calculated by means of (\ref{eq_prob}). For other pairs of configurations $W_{\{\vec{m}\},\{\vec{m}'\}}$ are equal to zero.

The periodicity of the system allows us also to take a Fourier transform of the master equation (\ref{master_eq}) to get
\begin{equation}
\frac{d}{dt}\mathbf{P}(\vec{k},t)=\hat{M}(\vec{k})\cdot\mathbf{P}(\vec{k},t),
\end{equation}
which is a matrix equation where the position vector $\vec{X}$ has been replaced by the wave vector $\vec{k}$. The components of the vector $\mathbf{P}(\vec{k},t)$ correspond to the configurations $\{\vec{m}\}$ and $\hat{M}(\vec{k})$ is the rate matrix, whose properties have been described in detail in Refs \onlinecite{Gortel,Zaluska-Kotur1,Zaluska-Kotur2}, with elements
\begin{equation}
M_{\{\vec{m}\},\{\vec{m}'\}}(\vec{k})=W_{\{\vec{m}\},\{\vec{m}'\}}(\vec{k})-\delta_{\{\vec{m}\},\{\vec{m}'\}}\sum_{\{\vec{m}''\}}W_{\{\vec{m}''\},\{\vec{m}\}},
\end{equation}
where $W_{\{\vec{m}\},\{\vec{m}'\}}(\vec{k})=F_{\{\vec{m}\},\{\vec{m}'\}}(\vec{k})W_{\{\vec{m}\},\{\vec{m}'\}}$ with $F_{\{\vec{m}\},\{\vec{m}'\}}(\vec{k})$ equal to 1 except when the reference particle crosses the boundary between elementary cells. In the latter case $F_{\{\vec{m}\},\{\vec{m}'\}}(\vec{k})=\exp(\pm i\vec{k}\vec{A})$, where $\vec{A}$ is the lattice vector between those cells and the sign depends on the direction of the jump.

When the system is brought out of the equilibrium state, it will start returning to that state with the rate described by the rate matrix's eigenvalues, which are always real and negative and correspond to the decay of the $k$th Fourier component of the fluctuation. The components of the left $\tilde{\mathbf{e}}(\vec{k})$ and the right $\mathbf{e}(\vec{k})$ eigenvector are related to each other by
\begin{equation}
e_{\{\vec{m}\}}(\vec{k})=P_{\{\vec{m}\}}^{eq}e_{\{\vec{m}\}}^{*}(\vec{k}).
\end{equation}

Close to the equilibrium the eigenvalue $\lambda_{D}(\vec{k})$ with the lowest absolute value is proportional to $k^2=\mid\vec{k}\mid^2$ and gives the diffusion tensor
\begin{equation}
\begin{bmatrix}
k_x&k_y
\end{bmatrix}
\begin{bmatrix}
D_{xx}&D_{xy}\\
D_{xy}&D_{yy}
\end{bmatrix}
\begin{bmatrix}
k_x\\
k_y
\end{bmatrix}
=-\lambda_{D}(\vec{k}).
\end{equation}
In order to find that eigenvalue we use a variational approach. We denote a trial eigenvector by $\tilde{\mathbf{\phi}}$ and we obtain its trial eigenvalue $\lambda_{D}^{var}(\vec{k})$ from
\begin{equation}
-\lambda_{D}^{var}(\vec{k})=-\lim_{k\rightarrow 0}\left(\frac{\tilde{\mathbf{\phi}}\cdot\hat{M}(\vec{k})\cdot\mathbf{\phi}}{\tilde{\mathbf{\phi}}\cdot\mathbf{\phi}}\right)=\lim_{k\rightarrow 0}\left(\frac{M(\vec{k})}{N(\vec{k})}\right),\label{var_form}
\end{equation}
which is the variational formula used by us to solve various diffusional problems \cite{Minkowski1,Gortel,Zaluska-Kotur1,Zaluska-Kotur2,Krzyzewski,Minkowski2,Minkowski3,Minkowski4,Minkowski5}. $M(\vec{k})$ and $N(\vec{k})$ have been referred to as the expectation value numerator and the normalization denominator, respectively. The most important step in the variational method is choosing the proper eigenvector. Following our earlier works we assume the form
\begin{equation}
\tilde{\phi}_{\{\vec{m}\}}^{*}=\sum_{i=0}^{N-1}\exp[i\vec{k}(\vec{m}_{i}+\vec{\delta}_{\vec{m}_{i}}+\vec{\Delta}_{\vec{m}_{i}})],\label{trial_vec}
\end{equation}
where $\vec{m}_{i}$ is the position of the $i$th particle in the $\{\vec{m}\}$ configuration, while $\vec{\delta}_{\vec{m}_{i}}$ and $\vec{\Delta}_{\vec{m}_{i}}$ are the variational parameters, called geometrical and correlational phase, respectively, assigned to that particle in that configuration. We insert that trial eigenvector into the variational formula (\ref{var_form}), which we then differentiate with respect to the variational parameters in order to find its minimal value.

The geometrical phase $\vec{\delta}_{\vec{m}_{i}}$ depends on the symmetry of the potential with respect to the site $\vec{m}_{i}$. By that we understand not only the nearest neighbors of that site but also the global structure of the potential. On the other hand, the correlational phase $\vec{\Delta}_{\vec{m}_{i}}$ is in general non-zero when two neighboring sites with different adsorption energies are both occupied. For a given particle the value of $\vec{\delta}_{\vec{m}_{i}}$ depends only on the position of that particle, while the value of $\vec{\Delta}_{\vec{m}_{i}}$ depends also on the occupancy of neighboring sites. Both for the geometrical and the correlational phases one can use symmetry arguments in order to deduce some properties of the values they can assume. For lattices exhibiting some kind of symmetry one can prove that some of the phases are equal to zero, or one can relate the value of one phase to that of another one.

Since $N(\vec{k})$ for our choice of the trial vector does not depend on the variational parameters, we shall consider the factors $M(\vec{k})$ and $N(\vec{k})$ separately. Therefore, in order to minimize the trial eigenvalue given by the formula (\ref{var_form}) it is enough to minimize its numerator $M(\vec{k})$. Following our earlier works we write it in the form
\begin{equation}
M(\vec{k})=\sum_{\{\vec{m}\},\{\vec{m}'\}}^{no\; rep}P_{\{\vec{m}'\}}^{eq}W_{\{\vec{m}\},\{\vec{m}'\}}\mid\tilde{\phi}_{\{\vec{m}'\}}^{*}(\vec{k})-\tilde{\phi}_{\{\vec{m}\}}^{*}(\vec{k})\mid^{2}.\label{numerator}
\end{equation}
Then we insert the trial vector (\ref{trial_vec}) into the above expression to find the values of the variational parameters for which the expression has the smallest possible value. In the limit $k\rightarrow 0$ and assuming that the configurations $\{\vec{m}\}$ and $\{\vec{m}'\}$ differ only by the position of one particle we can rewrite the numerator as
\begin{eqnarray}
M(\vec{k})&=&\sum_{A,B}^{no\;rep}W_{AB}\theta_{A}(1-\theta_{B})\sum_{\{n_{i}^{A}\}\cup\{n_{i}^{B}\}}P^{eq}(\{n_{i}^{A}\}\cup\{n_{i}^{B}\})\nonumber\\
&\times&[\vec{k}\cdot(\vec{a}_{AB}+\vec{\delta}_{B}-\vec{\delta}_{A}+\vec{\Delta}_{B}(\{n_{i}^{B}\})-\vec{\Delta}_{A}(\{n_{i}^{A}\}))]^{2},\nonumber\\
\label{numerator_2}
\end{eqnarray}
where the first summation is done over all the possible jumps in the system, that is all pairs of nearest neighbors with A denoted as the site from which the jump starts and B as the site to which the particle jumps. The second summation is performed over all possible occupational states of the neighbors of sites A and B. Because of the detailed balance condition (\ref{det_bal}) it is enough to consider each jump only in one direction, as indicated by the "no rep" comment over the sum symbol. $\{n_{i}^{A(B)}\}$ are sets of the occupation numbers of the neighbors of the A(B) site excluding the B(A) site. Sites A and B can in general have mutual neighbors, that is why in general $P^{eq}(\{n_{i}^{A}\}\cup\{n_{i}^{B}\})\neq P^{eq}(\{n_{i}^{A}\})P^{eq}(\{n_{i}^{B}\})$. As mentioned before, the values of the correlational phases depend on the occupancy of the neighboring sites. $\vec{a}_{AB}$ is the vector connecting the sites A and B.

On the other hand, the normalization denominator can be written as
\begin{equation}
N(\vec{k})=\sum_{\{\vec{m}\}}P_{\{\vec{m}\}}^{eq}\mid\tilde{\phi}_{\{\vec{m}\}}^{*}(\vec{k})\mid^{2},\label{denominator}
\end{equation}
which in the limit $k\rightarrow 0$ for particles with only exclusion interactions has a very simple form
\begin{equation}
N(0)=\sum_{i=0}^{N-1}\theta_{i}(1-\theta_{i}).
\end{equation}

Once we have the factors (\ref{numerator}) and (\ref{denominator}) expressed in terms of jump rates and equilibrium occupancies, the diffusion coefficient is the ratio of those factors.
\subsection{Numerator}
Calculating the numerator directly from the Eq. (\ref{numerator_2}) for a specific system is rather tedious as for each jump one has to consider all relevant configurations in the system, that is in general all occupational states of the neighborhood of the sites involved in the jump. However, we shall show now that we can perform the second sum in Eq. (\ref{numerator_2}) in a general case and obtain a formula which can be easily used for calculations. First of all, we assume that the total correlational phase of a particle is the sum of correlational phases related to all of its neighbors
\begin{equation}
\vec{\Delta}_{A}\{n_i^A\}=\sum_{i}n_{i}\vec{\Delta}_{Ai}.
\end{equation}
Then let us rewrite the second sum in Eq. (\ref{numerator_2}) as
\begin{eqnarray}
M_{AB}&=&\sum_{\{n_{i}^{A}\}\cup\{n_{i}^{B}\}}P^{eq}(\{n_{i}^{A}\}\cup\{n_{i}^{B}\})[\vec{k}\cdot(\vec{a}_{AB}+\vec{\delta}_{B}-\vec{\delta}_{A}\nonumber\\
&+&\vec{\Delta}_{B}(\{n_{i}^{B}\})-\vec{\Delta}_{A}(\{n_{i}^{A}\}))]^{2}=[\vec{k}\cdot(\vec{a}_{AB}+\vec{\delta}_{B}-\vec{\delta}_{A})]^{2}\nonumber\\
&+&\sum_{\{n_{i}^{A}\}\cup\{n_{i}^{B}\}}P^{eq}(\{n_{i}^{A}\}\cup\{n_{i}^{B}\})[\vec{k}\cdot(\vec{\Delta}_{B}-\vec{\Delta}_{A})]^2\nonumber\\
&+&2[\vec{k}\cdot(\vec{a}_{AB}+\vec{\delta}_{B}-\vec{\delta}_{A})]\sum_{\{n_{i}^{A}\}\cup\{n_{i}^{B}\}}P^{eq}(\{n_{i}^{A}\}\cup\{n_{i}^{B}\})\nonumber\\
&\times&[\vec{k}\cdot(\vec{\Delta}_{B}-\vec{\Delta}_{A})],\label{sum_AB}
\end{eqnarray}
where we used the fact that the geometrical phases do not depend on the occupancy of the neighboring sites so we could pull them out before the sum, which is equal to 1 if there are no correlational phases under it. In the above expression we have two sums. The one in the mixed term can be rewritten us
\begin{eqnarray}
\sum_{\{n_{i}^{A}\}\cup\{n_{i}^{B}\}}P^{eq}(\{n_{i}^{A}\}\cup\{n_{i}^{B}\})[\vec{k}\cdot(\vec{\Delta}_{B}-\vec{\Delta}_{A})]\\
=\sum_{i}^{\{B\}}\theta_{i}\vec{k}\cdot\vec{\Delta}_{Bi}-\sum_{i}^{\{A\}}\theta_{i}\vec{k}\cdot\vec{\Delta}_{Ai},\nonumber
\end{eqnarray}
where $\{A(B)\}$ denote the set of all neighbors of the site A(B) excluding the site B(A). On the other hand, the sum with the squared term is
\begin{eqnarray}
&&\sum_{\{n_{i}^{A}\}\cup\{n_{i}^{B}\}}P^{eq}(\{n_{i}^{A}\}\cup\{n_{i}^{B}\})[\vec{k}\cdot(\vec{\Delta}_{B}-\vec{\Delta}_{A})]^2\label{sum_squared}\\
&=&\sum_{\{n_{i}^{A}\}}P^{eq}(\{n_{i}^{A}\})[\vec{k}\cdot\vec{\Delta}_{A}]^2+\sum_{\{n_{i}^{B}\}}P^{eq}(\{n_{i}^{B}\})[\vec{k}\cdot\vec{\Delta}_{B}]^2\nonumber\\
&-&2\sum_{\{n_{i}^{A}\}\cup\{n_{i}^{B}\}}P^{eq}(\{n_{i}^{A}\}\cup\{n_{i}^{B}\})[\vec{k}\cdot\vec{\Delta}_{A}][\vec{k}\cdot\vec{\Delta}_{B}],\nonumber
\end{eqnarray}
where the first sum on the right-hand side can be written as
\begin{eqnarray}
&&\sum_{\{n_{i}^{A}\}}P^{eq}(\{n_{i}^{A}\})[\vec{k}\cdot\vec{\Delta}_{A}]^2=\sum_{i}^{\{A\}}\theta_{i}[\vec{k}\cdot\vec{\Delta}_{Ai}]^{2}\\
&+&\sum_{i\neq j}^{\{A\}}\theta_{i}\theta_{j}[\vec{k}\cdot\vec{\Delta}_{Ai}][\vec{k}\cdot\vec{\Delta}_{Aj}]=\sum_{i}^{\{A\}}\theta_{i}[\vec{k}\cdot\vec{\Delta}_{Ai}]^{2}\nonumber\\
&+&\Big\{\sum_{i}^{\{A\}}\theta_{i}[\vec{k}\cdot\vec{\Delta}_{Ai}]\Big\}^{2}-\sum_{i}^{\{A\}}\theta_{i}^{2}[\vec{k}\cdot\vec{\Delta}_{Ai}]^{2}\nonumber\\
&=&\Big\{\sum_{i}^{\{A\}}\theta_{i}[\vec{k}\cdot\vec{\Delta}_{Ai}]\Big\}^{2}+\sum_{i}^{\{A\}}\theta_{i}(1-\theta_{i})[\vec{k}\cdot\vec{\Delta}_{Ai}].\nonumber
\end{eqnarray}
and of course we can get an analogous expression for the second sum in (\ref{sum_squared}). Similarly one can show that the last sum there is
\begin{eqnarray}
&&\sum_{\{n_{i}^{A}\}\cup\{n_{i}^{B}\}}P^{eq}(\{n_{i}^{A}\}\cup\{n_{i}^{B}\})[\vec{k}\cdot\vec{\Delta}_{A}][\vec{k}\cdot\vec{\Delta}_{B}]\nonumber\\
&=&\sum_{i\neq j}^{\{A\}\cup\{B\}}\theta_{i}\theta_{j}[\vec{k}\cdot\vec{\Delta}_{Ai}][\vec{k}\cdot\vec{\Delta}_{Bj}]\\
&+&\sum_{i}^{\{A\}\cap\{B\}}\theta_{i}[\vec{k}\cdot\vec{\Delta}_{Ai}][\vec{k}\cdot\vec{\Delta}_{Bi}]\nonumber\\
&=&\sum_{i}^{\{A\}}\theta_{i}[\vec{k}\cdot\vec{\Delta}_{Ai}]\sum_{i}^{\{B\}}\theta_{i}[\vec{k}\cdot\vec{\Delta}_{Bi}]\nonumber\\
&+&\sum_{i}^{\{A\}\cap\{B\}}\theta_{i}(1-\theta_{i})[\vec{k}\cdot\vec{\Delta}_{Ai}][\vec{k}\cdot\vec{\Delta}_{Bi}].\nonumber
\end{eqnarray}
Now after using the results above and collecting all terms in the expression (\ref{sum_AB}) we can show that the final formula for the numerator in non-interacting case is
\begin{eqnarray}
M&=&\sum_{A,B}W_{AB}\theta_{A}(1-\theta_{B})\Big\{\Big[\vec{k}\cdot(\vec{a}_{AB}+\vec{\delta}_{B}-\vec{\delta}_{A})\nonumber\\
&+&\sum_{i}^{\{B\}}\theta_{i}\vec{k}\cdot\vec{\Delta}_{Bi}-\sum_{i}^{\{A\}}\theta_{i}\vec{k}\cdot\vec{\Delta}_{Ai}\Big]^2\label{var_form_nonint}\\
&+&\sum_{i}^{\{A\}\cup\{B\}}\theta_{i}(1-\theta_{i})[\vec{k}\cdot(\vec{\Delta}_{Bi}-\vec{\Delta}_{Ai})]^{2}\Big\}.\nonumber
\end{eqnarray}
After we assign variational parameters to the sites and to the pairs of neighbors in the system this formula is simple to use. In the next chapter we shall show a few examples illustrating how to use that formula to calculate the collective diffusion coefficient.
\section{Results}
We illustrate this method on several examples. Below we show coverage dependence of collective diffusion coefficients for lattices of complex energy landscapes calculated using variational approach. The formula (\ref{var_form_nonint}), which is the starting point of our calculations, depends on a few variational parameters, called geometrical and correlational phases, with respect to which it must be minimized. In general, one assigns one geometrical phase to each adsorption site and one correlational phase to each pair of nearest neighbors, however, usually it is possible to reduce the number of independent variational parameters using symmetry arguments.

In order to minimize the variational formula, one has to solve a set of linear equations obtained from differentiating it with respect to the assumed parameters. Such set of equations is always soluble, for example by means of Cramer's rule. Finally, the obtained values of the parameters are inserted into the formula (whose squared terms can be eliminated in the minimum, which simplifies the expression further) to get the diffusion coefficients as functions of jump rates, temperature and the system's coverage.

The above procedure is always the same and does not depend on the lattice symmetry. However, depending on the complexity of the investigated system calculations can be shorter or longer. We start with three examples of square lattices, and then two different hexagonal lattices are analyzed. Finally we show results for lattices that model experimental system of Ga diffusing on GaAs(001) surface. Diffusion coefficients for two different reconstructions of this surface are compared.

\subsection{Square lattice systems}
\subsubsection{Two equivalent sublattices}
\begin{figure}
	\includegraphics[width=8cm]{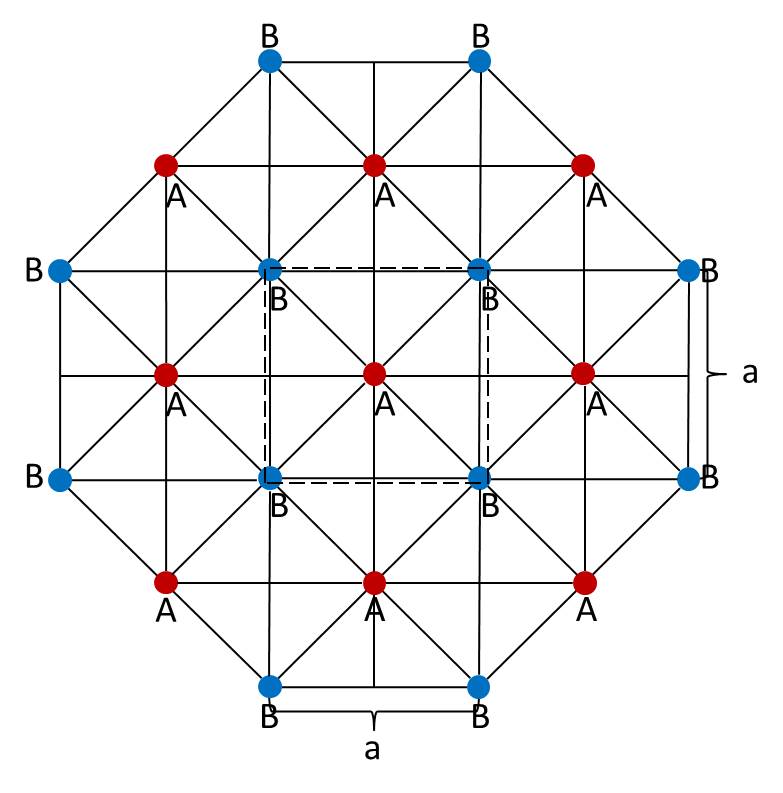}
	\caption{Lattice of square symmetry containing two different adsorption sites. Dashed line shows an elementary cell.}
	\label{2ads_sqr}
\end{figure}

The first  system  we  analyze is shown in Fig. \ref{2ads_sqr}. It is a lattice of square symmetry, similar to the checkered lattice considered in Ref. \onlinecite{Krzyzewski}, however, in addition to jumps between different sites there are also jumps between identical sites. One elementary cell contains one A and one B site. All jumps are symmetric and there are 8 possible jumps per elementary cell: 2 between A sites, 2 between B sites and 4 between A and B sites.

Because of the central inversion of the lattice in every adsorption site, all the geometrical phases are equal to zero. The only non-zero variational phases will be correlational ones related to the AB bonds. There will be 4 such phases in total, however, due to the lattice symmetry with respect to the rotation by 90 degrees, they will differ only by the direction. Therefore, they can be expressed by one variational parameter, $\vec{\Delta}_{AB}^{\pm\pm}=\Delta_{AB}(\pm 1;\pm 1)$, where the index $\pm\pm$ refers to the orientation of the AB bond at the lattice. The numerator of the variational formula is then
\begin{eqnarray}
M&=&\{W_{AA}\theta_{A}(1-\theta_{A})\left[a^2+16\theta_{B}(1-\theta_{B})\Delta_{AB}^{2}\right]\label{M_2ads_sqr}\\
&+&W_{BB}\theta_{B}(1-\theta_{B})\left[a^2+16\theta_{A}(1-\theta_{A})\Delta_{AB}^{2}\right]\nonumber\\
&+&4W_{AB}\theta_{A}(1-\theta_{B})\left[\left[\frac{a}{2}+\Delta_{AB}(\theta_{B}-\theta_{A})\right]^{2}\right.\nonumber\\
&+&\left.3\Delta_{AB}^{2}\left[\theta_{A}(1-\theta_{A})+\theta_{B}(1-\theta_{B})\right]\right]\}(k_x^2+k_y^2).\nonumber
\end{eqnarray}
After differentiating that formula with respect to $\Delta_{AB}$ and setting the derivative to zero we get
\begin{eqnarray}
\Delta_{AB}&=&W_{AB}(\theta_{A}-\theta_{B})a\Big/\Big\{2[4\theta_{B}(1-\theta_{A})(W_{AA}+W_{BB})\nonumber\\
&+&W_{AB}(\theta_{B}-\theta_{A})^{2}+3W_{AB}[\theta_{A}(1-\theta_{A})\label{Delta_2ads_sqr}\\
&+&\theta_{B}(1-\theta_{B})]]\Big\}.\nonumber
\end{eqnarray}
Using the condition for the derivative $\frac{\partial M}{\partial\Delta_{AB}}=0$ we can eliminate the square terms in the expression (\ref{M_2ads_sqr}) and
\begin{eqnarray}
M&=&a\{a\left[W_{AA}\theta_{A}(1-\theta_{A})+W_{BB}\theta_{B}(1-\theta_{B})\right.\nonumber\\
&+&\left.W_{AB}\theta_{A}(1-\theta_{B})\right]\label{M_2ads_sqr_simp}\\
&+&2W_{AB}\theta_{A}(1-\theta_{B})\Delta_{AB}(\theta_{B}-\theta_{A})\}(k_x^2+k_y^2).\nonumber
\end{eqnarray}
When phase (\ref{Delta_2ads_sqr}) is inserted into Eq. (\ref{M_2ads_sqr_simp}) we obtain the final expression for the numerator
\begin{eqnarray}
M&=&a^{2}\Big\{W_{AA}\theta_{A}(1-\theta_{A})+W_{BB}\theta_{B}(1-\theta_{B})\label{M_2ads_sqr_final}\\
&+&W_{AB}\theta_{A}(1-\theta_{B})-W_{AB}^{2}\theta_{A}(1-\theta_{B})(\theta_{B}-\theta_{A})^{2}\Big/\nonumber\\
&&\Big\{4\theta_{B}(1-\theta_{A})(W_{AA}+W_{BB})+W_{AB}(\theta_{B}-\theta_{A})^{2}\nonumber\\
&+&3W_{AB}\left[\theta_{A}(1-\theta_{A})+\theta_{B}(1-\theta_{B})\right]\Big\}\Big\}(k_{x}^{2}+k_{y}^{2}).\nonumber
\end{eqnarray}
Since there is only one adsorption site of each type in an elementary cell, the denominator is
\begin{equation}
N=\theta_{A}(1-\theta_{A})+\theta_{B}(1-\theta_{B}).\label{N_2ads_sqr}
\end{equation}
The diffusion is isotropic and $D_x=D_y=M/(N(k_x^2+k_y^2))$.

The first three terms in Eq. (\ref{M_2ads_sqr_final}), each of which contains one jump constant multiplied by adequate occupation probabilities, can be easily interpreted as contributions to the diffusion from respective paths without considering correlational effects. The term with $W_{AA}$ corresponds to diffusion through A sites, the one with $W_{BB}$ describes the diffusive path through B sites and the last one with $W_{AB}$ accounts for the diagonal path that goes alternately through A and B sites.

On the other hand, the last term in Eq. (\ref{M_2ads_sqr_final}) can be attributed to the inter-particle correlations in the system. It comes from the correlational phase that was initially assumed in the variational formula. That term is negative, which reflects the fact that the variational approach helps to estimate the upper value of the diffusion coefficient. Both in the limit $\theta\rightarrow 0$ (single particle) and $\theta\rightarrow 1$ (single hole) the correlation term vanishes. For a single particle the other terms together with the denominator (\ref{N_2ads_sqr}) yield the diffusion coefficient
\begin{figure}
	\includegraphics[width=6cm,angle=-90]{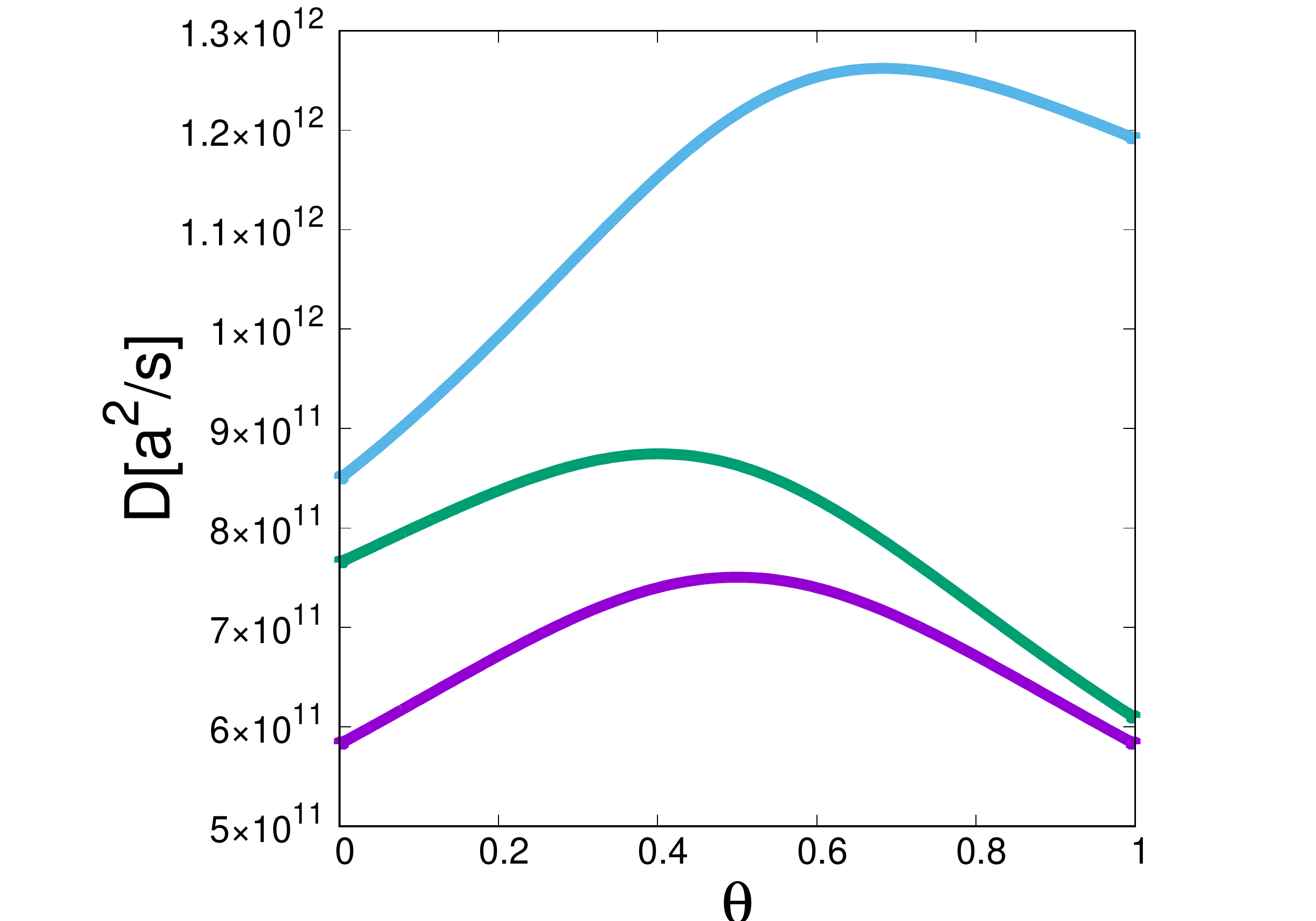}
	\caption{Dependence of the collective diffusion coefficient on the mean coverage of the system for the lattice shown in Fig. \ref{2ads_sqr}. For all curves $E_{A}$=0 eV, $E_{B}$=0.05 eV and $E_{AB}$=0.07 eV. For the bottom curve $E_{AA}=E_{BB}=\infty$, for the middle one $E_{AA}$=0.1 eV and $E_{BB}=\infty$, and for the upper curve $E_{AA}$=0.1 eV and $E_{BB}$=0.12 eV. Temperature T=300K and attempt frequency $\nu=10^{13}$/s.}
	\label{2ads_sqr_diff_th}
\end{figure}
\begin{equation}
D_p=\frac{W_{AA}W_{BA}+W_{BB}W_{AB}+W_{AB}W_{BA}}{W_{AB}+W_{BA}},
\end{equation}
which is identical to the expression derived in Ref. \onlinecite{Gosalvez1} for the same lattice. We used the detailed balance conditions for A and B sites $W_{AB}\theta_{A}(1-\theta_{B})=W_{BA}\theta_{B}(1-\theta_{A})$ while taking the single-particle limit. For a single hole the diffusion coefficient is
\begin{equation}
D_h=\frac{W_{AA}W_{AB}+W_{BB}W_{BA}+W_{AB}W_{BA}}{W_{AB}+W_{BA}}.
\end{equation}

For $W_{AA}=W_{BB}=0$ the diffusion at this lattice is the same as at the checkered lattice from Ref. \onlinecite{Krzyzewski}, $D_p=D_h$ and the maximum of the diffusion coefficient is for $\theta=0.5$. However, when we add jumps between sites of the same type, represented by $W_{AA}$ and $W_{BB}$, then the dependence of the diffusion coefficient on the coverage will change significantly. In Fig. \ref{2ads_sqr_diff_th} we show the dependence of the collective diffusion coefficient on the mean coverage of the system for three different cases. In all of them the depths of the adsorption sites are assumed $E_{A}=0$ eV and $E_{B}=0.05$ eV and the barrier between them is $E_{AB}=0.07$ eV. The barriers for $W_{AA}$ and $W_{BB}$ jumps differ between the curves. The bottom curve represents the diffusion where those barriers are infinite, that is the corresponding jumps are blocked. That case is equivalent to diffusion at the checkered lattice \cite{Krzyzewski}. The middle curve is for $E_{AA}=0.1$ eV and jumps between B sites are blocked. We can see that activating a channel between A sites speeds up the diffusion. Because A sites are deeper than B sites, the change in the diffusion coefficient is bigger for lower values of the concentration. Finally, the upper curve shows the diffusion for $E_{AA}=0.1$ eV and $E_{BB}=0.12$ eV. This time, as expected, the diffusion is more strongly accelerated for high values of $\theta$. Even though $E_{BB}>E_{AA}$, the effect is stronger than for the channel between A sites. That is because the value that decides about the jump rates is the difference between the barrier and the depth of the adsorption sites that are involved in the jump. In our case $E_{BB}-E_{B}<E_{AA}-E_{A}$.

\subsubsection{Lattice of square symmetry - inequivalent sublattices}
\begin{figure}
	\includegraphics[width=8cm]{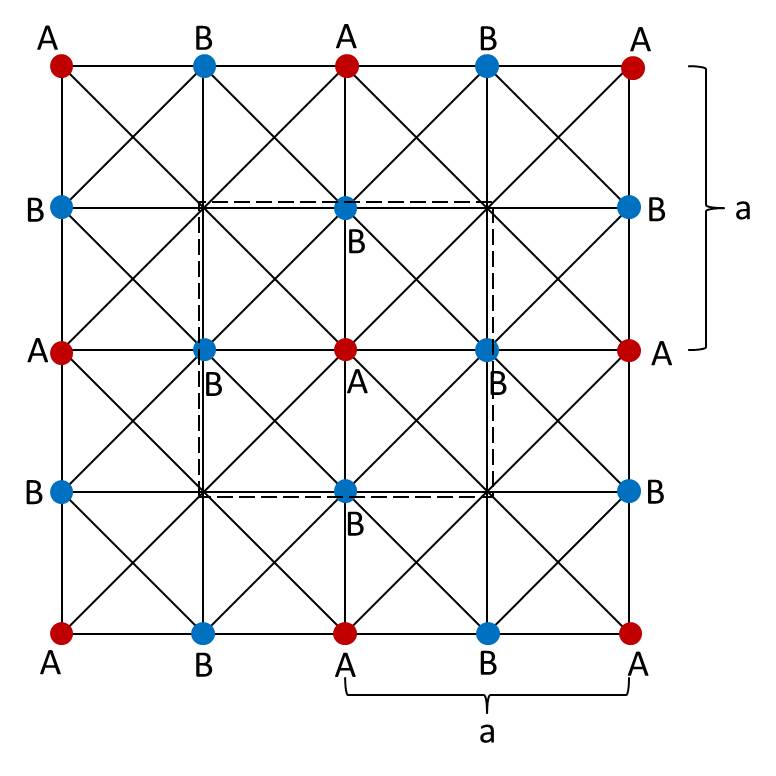}
	\caption{Lattice of square symmetry containing two different inequivalent adsorption sites. Dashed line shows an elementary cell.}
	\label{2ads_sqr2}
\end{figure}

Now let us consider a lattice shown in Fig. \ref{2ads_sqr2}. It possesses square symmetry but differs significantly from the square lattice in Fig. \ref{2ads_sqr}. In the geometric sense the adsorption sites A and B are not equivalent. Each A site connects to 4 other A sites and to 4  sites B, while each B site connects to 6  sites B and to 2  sites A. Each elementary cell contains one A site and, contrary to the previous examples, two B sites. Those two B sites differ by the angular orientation of the surrounding sites. They have the same adsorption energy but one of them is placed between the two A sites in the horizontal direction and the other in the vertical one. Therefore, the total number of sites per elementary cell is 3. One should also note that there are two types of jumps between the B sites. One of them is a short jump in the diagonal direction and the other one is a long jump in the horizontal or vertical direction. We will denote the corresponding jump rates by $W_{BB}^{s}$ and $W_{BB}^{l}$, respectively.

Again, there will be one correlation phase, which will possess four versions with respect to the angular orientation. As in the previous case they can all be expressed by one variational parameter $\Delta_{AB}$. The numerator is
\begin{eqnarray}
M&=&\Big\{2W_{AA}\theta_{A}(1-\theta_{B})\Big\{\left[\frac{a}{2}+(\theta_{B}-\theta_{A})\Delta_{AB}\right]^{2}\label{M_2ads_sqr_2}\\
&+&\Delta_{AB}^{2}\left[\theta_{A}(1-\theta_{A})+3\theta_{B}(1-\theta_{B})\right]\Big\}\nonumber\\
&+&2W_{AA}\theta_{A}(1-\theta_{A})\left[a^{2}+4\theta_{B}(1-\theta_{B})\Delta_{AB}^{2}\right]\nonumber\\
&+&4W_{BB}^{s}\theta_{B}(1-\theta_{B})\left[\left(\frac{a}{2}\right)^{2}+2\theta_{A}(1-\theta_{A})\Delta_{AB}^{2}\right]\nonumber\\
&+&W_{BB}^{l}\theta_{B}(1-\theta_{B})\left[a^{2}+4\theta_{A}(1-\theta_{A})\Delta_{AB}^{2}\right]\Big\}(k_{x}^{2}+k_{y}^{2}).\nonumber
\end{eqnarray}
From there we obtain
\begin{eqnarray}
\Delta_{AB}&=&W_{AB}(\theta_{A}-\theta_{B})a\Big/\Big\{2\big\{W_{AB}\left[(\theta_{B}-\theta_{A})^{2}+\theta_{A}(1-\theta_{A})\right.\nonumber\\
&+&\left.3\theta_{B}(1-\theta_{B})\right]+2(1-\theta_{A})\theta_{B}\left[2(W_{AA}+W_{BB}^{s})\right.\\
&+&\left.W_{BB}^{l}\right]\big\}\Big\}\nonumber
\end{eqnarray}
and
\begin{eqnarray}
M&=&a^{2}\Big\{2W_{AA}\theta_{A}(1-\theta_{A})+(W_{BB}^{s}+W_{BB}^{l})\theta_{B}(1-\theta_{B})\nonumber\\
&+&\frac{1}{2}W_{AB}\theta_{A}(1-\theta_{B})-W_{AB}^{2}\theta_{A}(1-\theta_{B})(\theta_{B}-\theta_{A})^{2}\Big/\nonumber\\
&&\Big\{2\big\{W_{AB}\left[(\theta_{B}-\theta_{A})^{2}+\theta_{A}(1-\theta_{A})+3\theta_{B}(1-\theta_{B})\right]\nonumber\\
&+&2(1-\theta_{A})\theta_{B}\left[2(W_{AA}+W_{BB}^{s})+W_{BB}^{l}\right]\big\}\Big\}\Big\}(k_{x}^{2}+k_{y}^{2}).\nonumber\\
\end{eqnarray}
The denominator in this case is
\begin{equation}
N=\theta_{A}(1-\theta_{A})+2\theta_{B}(1-\theta_{B}).
\end{equation}

\begin{figure}
	\includegraphics[width=6cm,angle=-90]{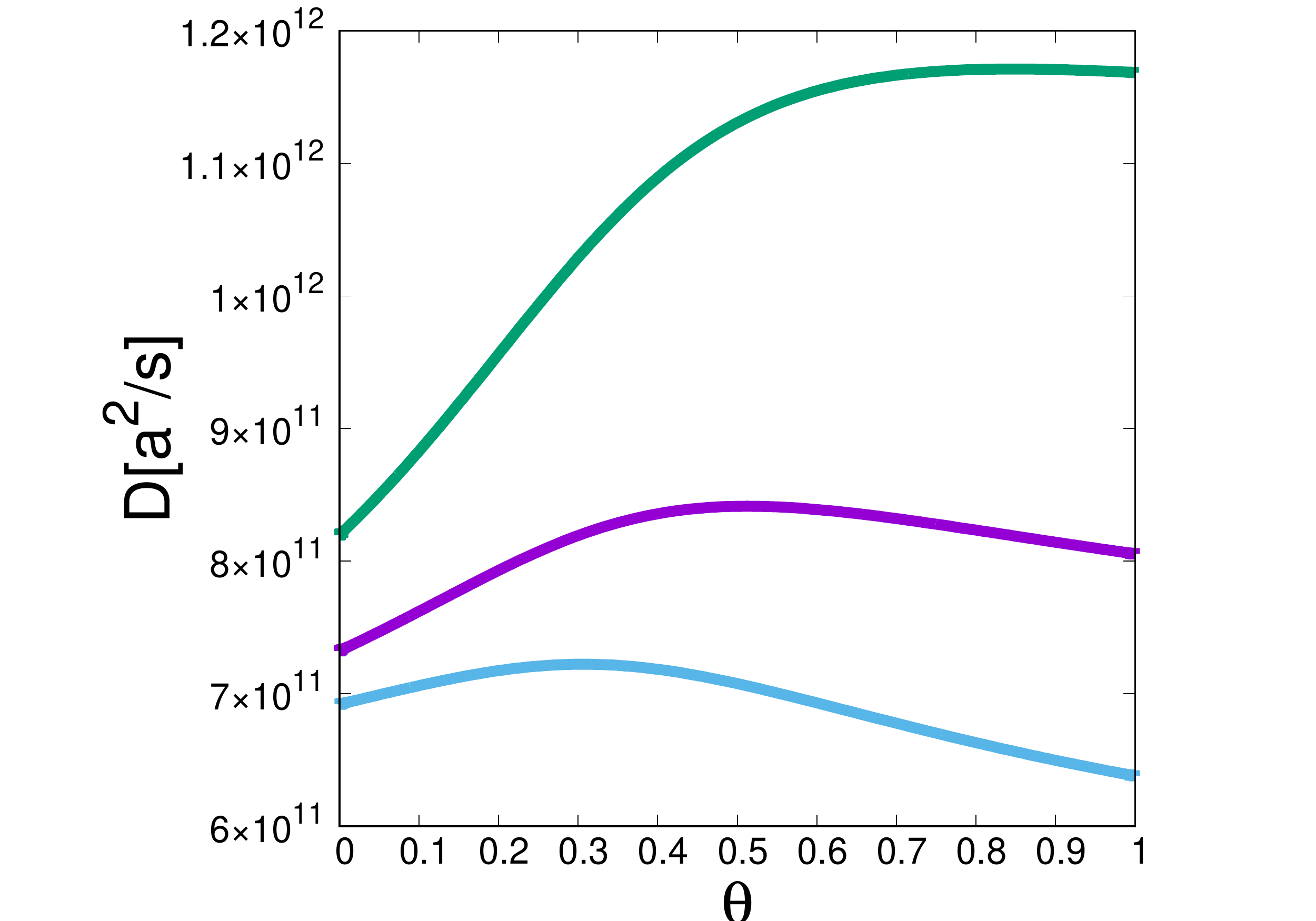}
	\caption{Dependence of the collective diffusion coefficient on the mean coverage of the system for the lattice shown in Fig. \ref{2ads_sqr2}. For all curves $E_{A}$=0 eV, $E_{B}$=0.05 eV, $E_{AA}$=0.1 eV, $E_{BB}^{s}$=0.12 eV and $E_{AB}$=0.07 eV. For the bottom curve $E_{BB}^{l}$=0.14 eV, for the middle one $E_{BB}^{l}$=0.12 eV and for the upper curve $E_{BB}^{l}$=0.1 eV. Temperature T=300K and attempt frequency $\nu=10^{13}$/s.}
	\label{2ads_sqr2_diff_th}
\end{figure}

The diffusion coefficient is calculated again by $D_x=D_y=M/(N(k_x^2+k_y^2))$ and its dependence on the mean coverage for three sets of the jump rates is shown in Fig. \ref{2ads_sqr2_diff_th}. As can be expected when we decrease energy barrier for long distance-jumps  between $B$ and  $B$ sites $E^{l}_{BB}$, diffusion coefficients increase for all surface coverages. However, interestingly the increase of high-coverage diffusion coefficient is much faster. As a result for higher barrier low-coverage diffusion coefficient is higher than high-coverage diffusion coefficient, while for lower barrier high-coverage diffusion becomes higher by three orders of magnitude than the low-coverage one.    

\subsubsection{Square lattice with three different adsorption sites}
\begin{figure}
	\includegraphics[width=8cm]{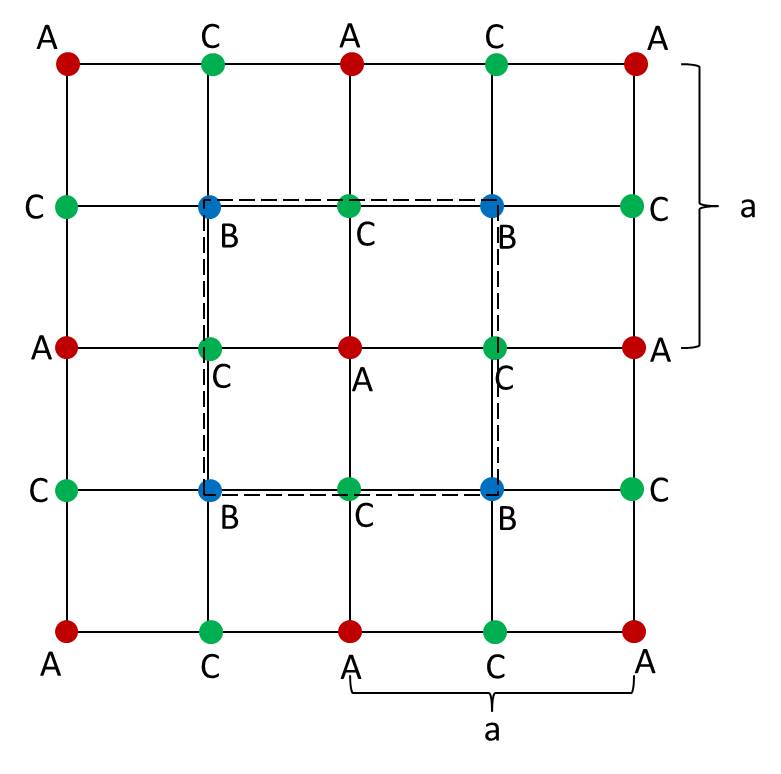}
	\caption{Lattice of square symmetry containing three different adsorption sites. Dashed line shows an elementary cell.}
	\label{3ads_sqr}
\end{figure}

We analyze diffusion at the lattice shown in Fig. \ref{3ads_sqr}. The lattice is of square symmetry with three different adsorption sites. In general they all differ by the adsorption energy. Sites A and B are geometrically equivalent to each other in the sense that they both have four C neighbors, one at each of the four main directions. Sites C, on the other hand, have two A neighbors placed either horizontally or vertically and two B neighbors placed in the perpendicular direction. Each elementary cell has one A site, one B site and two C sites. The two C sites differ by the angular orientation of their neighbors.

There are two types of jumps between adsorption sites: $W_{AC}$ and $W_{BC}$. Per elementary cell there are four $W_{AC}$ jumps and four $W_{BC}$ jumps. Again, because of the symmetry of the adsorption sites there are no geometrical phases. There are two correlational phases, $\Delta_{AC}$ and $\Delta_{BC}$, related to the respective pairs of neighbours. The numerator is
\begin{eqnarray}
M&=&\Big\{2W_{AC}\theta_{A}(1-\theta_{C})\Big\{\left[\frac{a}{2}+\Delta_{AC}(\theta_{C}-\theta_{A})\right]^{2}\\
&+&\Delta_{AC}^{2}\left[\theta_{A}(1-\theta_{A})+3\theta_{C}(1-\theta_{C})\right]+2\Delta_{BC}^{2}\theta_{B}(1-\theta_{B})\Big\}\nonumber\\
&+&2W_{BC}\theta_{B}(1-\theta_{C})\Big\{\left[\frac{a}{2}+\Delta_{BC}(\theta_{C}-\theta_{B})\right]^{2}\nonumber\\
&+&\Delta_{BC}^{2}\left[\theta_{B}(1-\theta_{B})+3\theta_{C}(1-\theta_{C})\right]+2\Delta_{AC}^{2}\theta_{A}(1-\theta_{A})\Big\}\Big\}\nonumber\\
&\times&(k_x^2+k_y^2).\nonumber
\end{eqnarray}
As before, in order to find the variational parameters we have to differentiate the above expression with respect to them and set the derivatives to zero. That will give us a set of two equations. In the case considered here each equation will contain only one phase so one can easily calculate their values
\begin{eqnarray}
\Delta_{AC}&=&aW_{AC}(\theta_{A}-\theta_{C})\Big/\{2\{W_{AC}[(\theta_{C}-\theta_{A})^{2}+\theta_{A}(1-\theta_{A})\nonumber\\
&+&3\theta_{C}(1-\theta_{C})]+2W_{BC}\theta_{B}(1-\theta_{A})\}\},\\
\Delta_{BC}&=&aW_{BC}(\theta_{B}-\theta_{C})\Big/\{2\{W_{BC}[(\theta_{C}-\theta_{B})^{2}+\theta_{B}(1-\theta_{B})\nonumber\\
&+&3\theta_{C}(1-\theta_{C})]+2W_{AC}\theta_{A}(1-\theta_{B})\}\}.\nonumber
\end{eqnarray}
Finally we get the numerator
\begin{eqnarray}
M&=&\frac{a^{2}}{2}\Big\{W_{AC}\theta_{A}(1-\theta_{C})+W_{BC}\theta_{B}(1-\theta_{C})\label{M_3ads_sqr_final}\\
&-&W_{AC}^{2}\theta_{A}(1-\theta_{C})(\theta_{C}-\theta_{A})^{2}\Big/\Big\{W_{AC}[(\theta_{C}-\theta_{A})^{2}\nonumber\\
&+&\theta_{A}(1-\theta_{A})+3\theta_{C}(1-\theta_{C})]+2W_{BC}\theta_{B}(1-\theta_{A})\Big\}\nonumber\\
&-&W_{BC}^{2}\theta_{B}(1-\theta_{C})(\theta_{C}-\theta_{B})^{2}\Big/\Big\{W_{BC}[(\theta_{C}-\theta_{B})^{2}\nonumber\\
&+&\theta_{B}(1-\theta_{B})+3\theta_{C}(1-\theta_{C})]+2W_{AC}\theta_{A}(1-\theta_{B})\Big\}\Big\}\nonumber\\
&\times&(k_x^2+k_y^2).
\end{eqnarray}
The denominator is $N=\theta_{A}(1-\theta_{A})+\theta_{B}(1-\theta_{B})+2\theta_{C}(1-\theta_{C})$.

From the above expressions one can simply calculate the diffusion coefficient as the ratio $D=M/(N(k_x^2+k_y^2))$. From Eq. (\ref{M_3ads_sqr_final}) it is seen that there are two main diffusive paths: one through $W_{AC}$ jumps and the other through $W_{BC}$ jumps. Again, the negative terms are due to the inter-particle correlations.
\subsection{Hexagonal lattice systems}
\begin{figure}
	\includegraphics[width=8cm]{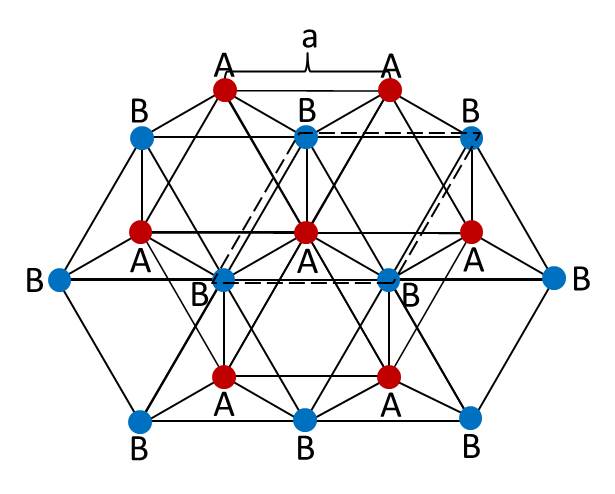}
	\caption{Lattice of hexagonal symmetry containing two different adsorption sites. Dashed line shows an elementary cell.}
	\label{2ads_hex}
\end{figure}
\subsubsection{System of two hexagonal sublattices}
Next we analyze a lattice of hexagonal symmetry shown in Fig. \ref{2ads_hex}. Here each site contains 6 connections to sites of the same type and 3 connections to sites of the other type.

Similarly to the previous cases, using symmetry arguments one can reduce the number of independent variational phases to one correlation phase between A and B sites. The numerator of the diffusion coefficient is
\begin{eqnarray}
M&=&\{\frac{3}{2}W_{AA}\theta_{A}(1-\theta_{A})\left[a^2+6\theta_{B}(1-\theta_{B})\Delta_{AB}^{2}\right]\label{M_2ads_hex}\\
&+&\frac{3}{2}W_{BB}\theta_{B}(1-\theta_{B})\left[a^2+6\theta_{A}(1-\theta_{A})\Delta_{AB}^{2}\right]\nonumber\\
&+&\frac{3}{2}W_{AB}\theta_{A}(1-\theta_{B})\left[\left[\frac{\sqrt{3}}{2}a+\Delta_{AB}(\theta_{B}-\theta_{A})\right]^{2}\right.\nonumber\\
&+&\left.2\Delta_{AB}^{2}\left[\theta_{A}(1-\theta_{A})+\theta_{B}(1-\theta_{B})\right]\right]\}(k_x^2+k_y^2).\nonumber
\end{eqnarray}
From there one can find the correlational phase
\begin{eqnarray}
\Delta_{AB}&=&\sqrt{3}W_{AB}(\theta_{A}-\theta_{B})a\Big/\Big\{3[6\theta_{B}(1-\theta_{A})(W_{AA}+W_{BB})\nonumber\\
&+&W_{AB}(\theta_{B}-\theta_{A})^{2}+2W_{AB}[\theta_{A}(1-\theta_{A})\\
&+&\theta_{B}(1-\theta_{B})]]\Big\},\nonumber
\end{eqnarray}
which gives the final value of the numerator
\begin{eqnarray}
M&=&\frac{a^{2}}{2}\Big\{3W_{AA}\theta_{A}(1-\theta_{A})+3W_{BB}\theta_{B}(1-\theta_{B})\label{M_2ads_hex_final}\\
&+&W_{AB}\theta_{A}(1-\theta_{B})-W_{AB}^{2}\theta_{A}(1-\theta_{B})(\theta_{B}-\theta_{A})^{2}\Big/\nonumber\\
&&\Big\{6\theta_{B}(1-\theta_{A})(W_{AA}+W_{BB})+W_{AB}(\theta_{B}-\theta_{A})^{2}\nonumber\\
&+&2W_{AB}\left[\theta_{A}(1-\theta_{A})+\theta_{B}(1-\theta_{B})\right]\Big\}\Big\}(k_{x}^{2}+k_{y}^{2}).\nonumber
\end{eqnarray}
The denominator is the same as for the case of two square sublattices, that is $N=\theta_{A}(1-\theta_{A})+\theta_{B}(1-\theta_{B})$ and again $D_x=D_y=M/(N(k_x^2+k_y^2))$.

Expression (\ref{M_2ads_hex_final}) has in fact the same structure as the numerator (\ref{M_2ads_sqr_final}) for the square lattice. The only difference is in the coefficients that stand by various terms. That is not surprising since both lattices contain two different adsorption sites per elementary cell and they are connected with each other similarly at both lattices. They differ by the number and the geometry of those connections but at both lattices the three main diffusive paths are along the $W_{AA}$, $W_{BB}$ and $W_{AB}$ jumps.

\begin{figure}
	\includegraphics[width=6cm,angle=-90]{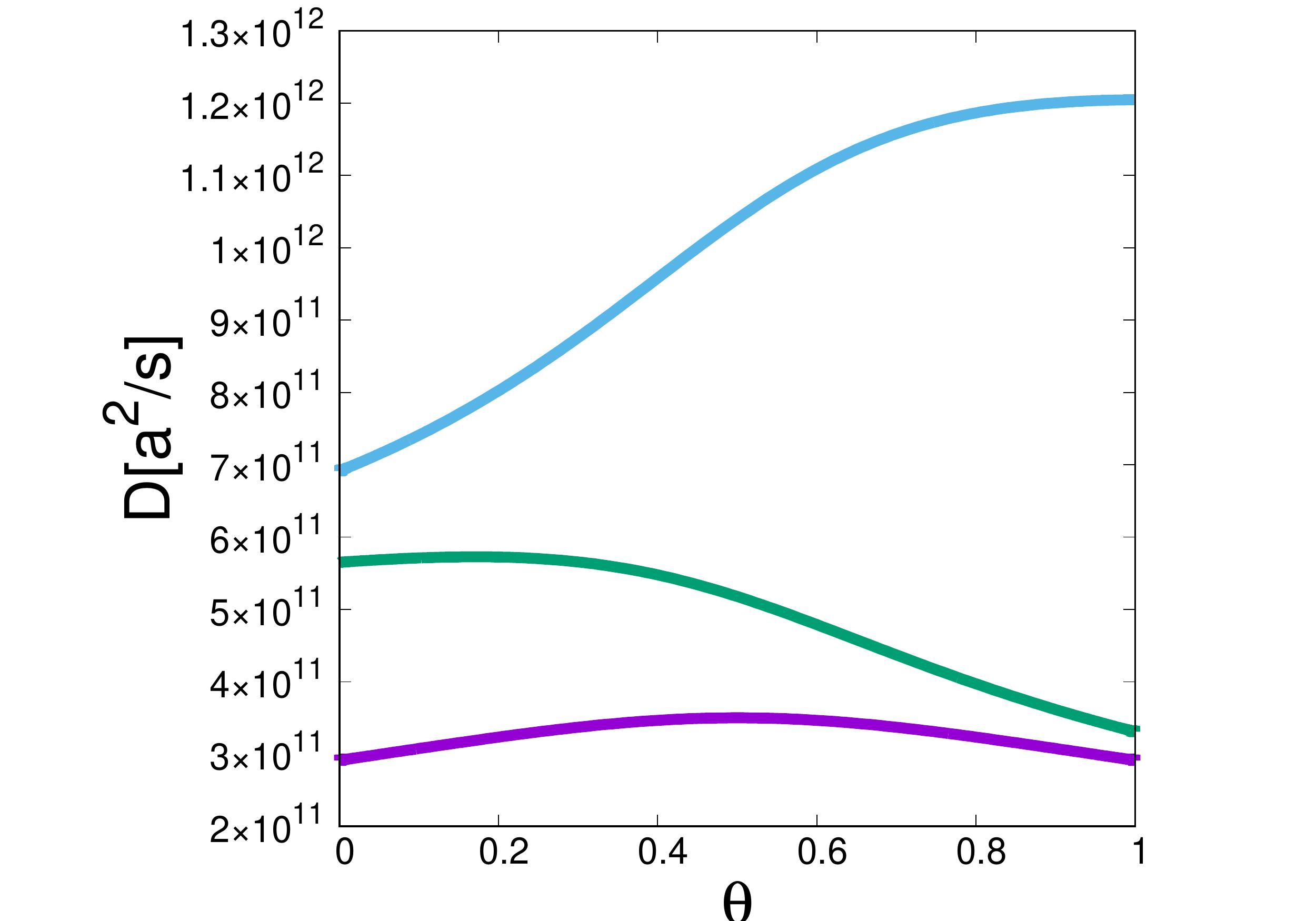}
	\caption{Dependence of the collective diffusion coefficient on the mean coverage of the system for the lattice shown in Fig. \ref{2ads_hex}. For all curves $E_{A}$=0 eV, $E_{B}$=0.05 eV and $E_{AB}$=0.07 eV. For the bottom curve $E_{AA}=E_{BB}=\infty$, for the middle one $E_{AA}$=0.1 eV and $E_{BB}=\infty$, and for the upper curve $E_{AA}$=0.1 eV and $E_{BB}$=0.12 eV. Temperature T=300K and attempt frequency $\nu=10^{13}$/s.}
	\label{2ads_hex_diff_th}
\end{figure}

We show the collective diffusion coefficient for the hexagonal lattice in Fig. \ref{2ads_hex_diff_th} for the same set of parameters as for the square lattice from  Figs \ref{2ads_sqr} and \ref{2ads_sqr_diff_th}. The behavior of the diffusion is qualitatively the same in both cases but we can see that here the diffusion is generally lower and the curves are flatter than in Fig. \ref{2ads_sqr_diff_th}, especially the one for $E_{AA}=E_{BB}=\infty$.

\begin{figure}
	\includegraphics[width=8cm]{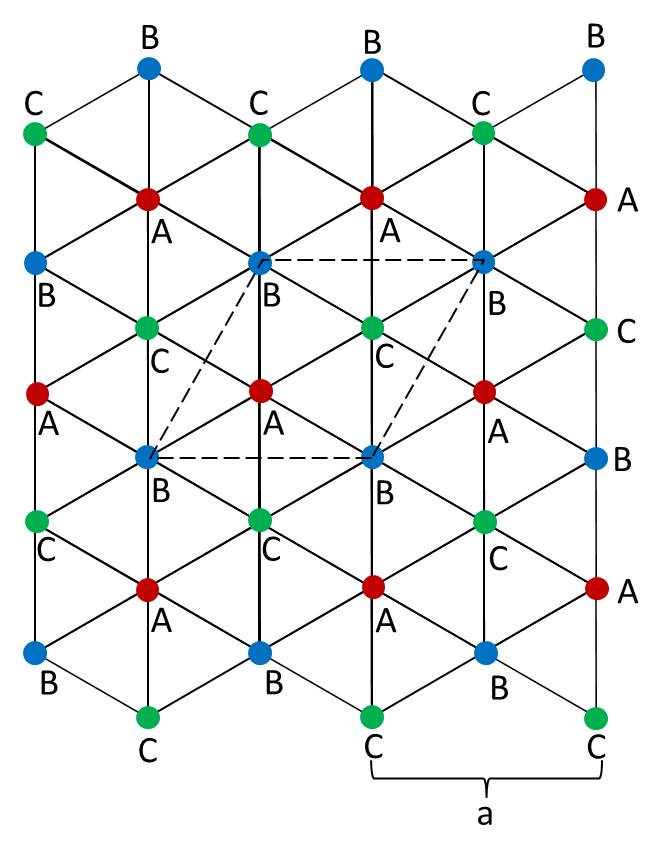}
	\caption{Lattice of hexagonal symmetry containing three different adsorption sites. Dashed line shows an elementary cell.}
	\label{3ads_hex}
\end{figure}

\subsubsection{Lattice of hexagonal symmetry - three sublattices}
In Fig. \ref{3ads_hex} we show another lattice with three different adsorption sites. All of them have the same geometry, each site of a given type has six connections to sites of two other types in that way that A site connects to 3 B and 3 C sites and so on.

There are three possible types of jumps in the system: $W_{AB}$, $W_{BC}$ and $W_{CA}$ and per elementary cell each of those jumps exists in three angular versions. Contrary to the previous examples, in this case there are three independent correlational phases: $\Delta_{AB}$, $\Delta_{BC}$ and $\Delta_{CA}$. The numerator is
\begin{eqnarray}
M&=&\Bigg\{3W_{AB}\theta_{A}(1-\theta_{B})\left[\frac{1}{2}\left[\frac{\sqrt{3}}{3}a+\Delta_{AB}(\theta_{B}-\theta_{A})\right]^{2}\right.\nonumber\\
&+&[\theta_{A}(1-\theta_{A})+\theta_{B}(1-\theta_{B})]\Delta_{AB}^{2}\nonumber\\
&+&\left.\theta_{C}(1-\theta_{C})\left[\frac{3}{2}(\Delta_{BC}^{2}+\Delta_{CA}^{2})+\Delta_{BC}\Delta_{CA}\right]\right]\nonumber\\
&+&3W_{BC}\theta_{B}(1-\theta_{C})\left[\frac{1}{2}\left[\frac{\sqrt{3}}{3}a+\Delta_{BC}(\theta_{C}-\theta_{B})\right]^{2}\right.\nonumber\\
&+&[\theta_{B}(1-\theta_{B})+\theta_{C}(1-\theta_{C})]\Delta_{BC}^{2}\nonumber\\
&+&\left.\theta_{A}(1-\theta_{A})\left[\frac{3}{2}(\Delta_{CA}^{2}+\Delta_{AB}^{2})+\Delta_{CA}\Delta_{AB}\right]\right]\nonumber\\
&+&3W_{CA}\theta_{C}(1-\theta_{A})\left[\frac{1}{2}\left[\frac{\sqrt{3}}{3}a+\Delta_{CA}(\theta_{A}-\theta_{C})\right]^{2}\right.\nonumber\\
&+&[\theta_{C}(1-\theta_{C})+\theta_{A}(1-\theta_{A})]\Delta_{CA}^{2}\nonumber\\
&+&\left.\theta_{B}(1-\theta_{B})\left[\frac{3}{2}(\Delta_{AB}^{2}+\Delta_{BC}^{2})+\Delta_{AB}\Delta_{BC}\right]\right]\Bigg\}\nonumber\\
&\times&(k_x^2+k_y^2).\label{M_3ads_hex}
\end{eqnarray}
Unlike previous examples, here mixed terms (for example $\Delta{AB}\Delta{BC}$) enter the expression for the numerator. That means that the derivatives will contain also other phases than that with respect to which the expression is differentiated. We shall write the set of linear equations for the correlational phases in the matrix form
\begin{equation}
\begin{bmatrix}
X_{11}&X_{12}&X_{13}\\
X_{21}&X_{22}&X_{23}\\
X_{31}&X_{32}&X_{33}\\
\end{bmatrix}
\begin{bmatrix}
\Delta_{AB}\\
\Delta_{BC}\\
\Delta_{CA}
\end{bmatrix}
=
\begin{bmatrix}
Y_{1}\\
Y_{2}\\
Y_{3}
\end{bmatrix}
\label{set_3eq}
\end{equation}
where
\begin{eqnarray}
X_{11}&=&W_{AB}\theta_{A}(1-\theta_{B})\left[(\theta_{B}-\theta_{A})^{2}+2\left[\theta_{A}(1-\theta_{A})\right.\right.\nonumber\\
&+&\left.\left.\theta_{B}(1-\theta_{B})\right]\right]+3\theta_{B}(1-\theta_{A})(W_{BC}\theta_{A}(1-\theta_{C})\nonumber\\
&+&W_{CA}\theta_{C}(1-\theta_{B}))\nonumber\\
X_{22}&=&W_{BC}\theta_{B}(1-\theta_{C})\left[(\theta_{C}-\theta_{B})^{2}+2\left[\theta_{B}(1-\theta_{B})\right.\right.\nonumber\\
&+&\left.\left.\theta_{C}(1-\theta_{C})\right]\right]+3\theta_{C}(1-\theta_{B})(W_{CA}\theta_{B}(1-\theta_{A})\nonumber\\
&+&W_{AB}\theta_{A}(1-\theta_{C}))\nonumber\\
X_{33}&=&W_{CA}\theta_{C}(1-\theta_{A})\left[(\theta_{A}-\theta_{C})^{2}+2\left[\theta_{C}(1-\theta_{C})\right.\right.\nonumber\\
&+&\left.\left.\theta_{A}(1-\theta_{A})\right]\right]+3\theta_{A}(1-\theta_{C})(W_{AB}\theta_{C}(1-\theta_{B})\nonumber\\
&+&W_{BC}\theta_{B}(1-\theta_{A}))\nonumber\\
X_{12}&=&X_{21}=W_{CA}\theta_{C}(1-\theta_{A})\theta_{B}(1-\theta_{B})\nonumber\\
X_{13}&=&X_{31}=W_{BC}\theta_{B}(1-\theta_{C})\theta_{A}(1-\theta_{A})\nonumber\\
X_{23}&=&X_{32}=W_{AB}\theta_{A}(1-\theta_{B})\theta_{C}(1-\theta_{C})\nonumber\\
Y_{1}&=&-W_{AB}\theta_{A}(1-\theta_{B})(\theta_{B}-\theta_{A})\frac{\sqrt{3}}{3}a\nonumber\\
Y_{2}&=&-W_{BC}\theta_{B}(1-\theta_{C})(\theta_{C}-\theta_{B})\frac{\sqrt{3}}{3}a\nonumber\\
Y_{3}&=&-W_{CA}\theta_{C}(1-\theta_{A})(\theta_{A}-\theta_{C})\frac{\sqrt{3}}{3}a.
\end{eqnarray}
As we can see this case is more complex than the previous ones and also it is more difficult to obtain the solution. However, one can solve the set of equations using Cramer's rule
\begin{eqnarray}
\Delta_{AB}&=&\frac{\det\begin{bmatrix}
	Y_{1}&X_{12}&X_{13}\\
	Y_{2}&X_{22}&X_{23}\\
	Y_{3}&X_{32}&X_{33}\\
	\end{bmatrix}}{\det\begin{bmatrix}
	X_{11}&X_{12}&X_{13}\\
	X_{21}&X_{22}&X_{23}\\
	X_{31}&X_{32}&X_{33}\\
	\end{bmatrix}}\nonumber\\
\Delta_{BC}&=&\frac{\det\begin{bmatrix}
	X_{11}&Y_{1}&X_{13}\\
	X_{21}&Y_{2}&X_{23}\\
	X_{31}&Y_{3}&X_{33}\\
	\end{bmatrix}}{\det\begin{bmatrix}
	X_{11}&X_{12}&X_{13}\\
	X_{21}&X_{22}&X_{23}\\
	X_{31}&X_{32}&X_{33}\\
	\end{bmatrix}}\label{3ads_hex_phases}\\
\Delta_{CA}&=&\frac{\det\begin{bmatrix}
	X_{11}&X_{12}&Y_{1}\\
	X_{21}&X_{22}&Y_{2}\\
	X_{31}&X_{32}&Y_{3}\\
	\end{bmatrix}}{\det\begin{bmatrix}
	X_{11}&X_{12}&X_{13}\\
	X_{21}&X_{22}&X_{23}\\
	X_{31}&X_{32}&X_{33}\\
	\end{bmatrix}}.\nonumber
\end{eqnarray}
Again, we can simplify the expression (\ref{M_3ads_hex}) by using the condition of vanishing derivatives to get
\begin{eqnarray}
M&=&\frac{a^2}{2}\Big\{W_{AB}\theta_{A}(1-\theta_{B})+W_{BC}\theta_{B}(1-\theta_{C})\\
&+&W_{CA}\theta_{C}(1-\theta_{A})\Big\}\nonumber\\
&+&\frac{\sqrt{3}}{2}a\Big\{W_{AB}\theta_{A}(1-\theta_{B})(\theta_{B}-\theta_{A})\Delta_{AB}\nonumber\\
&+&W_{BC}\theta_{B}(1-\theta_{C})(\theta_{C}-\theta_{B})\Delta_{BC}\nonumber\\
&+&W_{CA}\theta_{C}(1-\theta_{A})(\theta_{A}-\theta_{C})\Delta_{CA}\Big\}\nonumber\\
&\times&(k_x^2+k_y^2).\nonumber
\end{eqnarray}
Then we put the phases (\ref{3ads_hex_phases}) into the above expression and after dividing it by the denominator
\begin{equation}
N=\theta_{A}(1-\theta_{A})+\theta_{B}(1-\theta_{B})+\theta_{C}(1-\theta_{C})
\end{equation}
we get the collective diffusion coefficient $D$.

\subsection{Diffusion on GaAs(001)surface}
The variational method is not limited to lattices whose adsorption sites possess symmetry. It can be just as well used to analyze diffusion in systems where there is an asymmetry of the jumps from a given site. In such cases one should assume a geometrical phase related to that site. In general, one should look not only at the symmetry of the jumps from that site but also at the symmetry of the entire system. Even when the jumps are symmetric, the geometrical phase can be non-zero when the whole potential is not symmetric with respect to the site considered. The whole procedure of finding the diffusion coefficients is the same as for symmetric landscapes. The variational formula is minimized with respect to the parameters by simple differentiation, which gives a set of linear equations.

In this section we shall consider two examples of lattices which contain at least one site with respect to which the energetic landscape is asymmetric and consequently its geometrical phase cannot be set to zero. These examples model Ga adatoms diffusion on GaAs(001) surfaces in two different reconstructions. We show how diffusion anisotropy changes with the system coverage in both cases.
\subsubsection{Ga adatoms on GaAs(001)-c(4x4)}
\begin{figure}
\includegraphics[width=8cm]{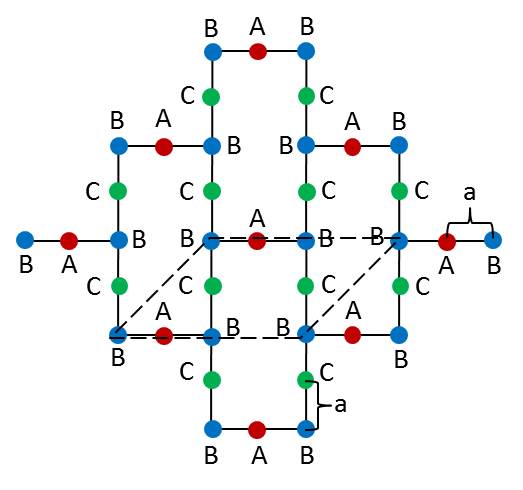}
\caption{Lattice for Ga adatom diffusing at GaAs(001)-c(4x4), based on Ref. \onlinecite{LePage}. Dashed line shows an elementary cell.}
\label{GaAs4x4}
\end{figure}

In Fig. \ref{GaAs4x4} a simplified lattice for Ga adatom at reconstructed GaAs(001)-c(4x4) surface is shown. The scheme of jumps is based on ab-initio calculations for that system in Ref. \onlinecite{LePage}. Based on those data diffusion coefficients for a single Ga atom were calculated in Ref. \onlinecite{Gosalvez2}. A more detailed landscape was found in Ref. \onlinecite{Roehl2}, which contained a few more adsorption sites and energetic barriers. Single-particle diffusion coefficients based on those new data were found in Ref. \onlinecite{Minkowski2} by means of our variational approach.

Here we shall calculate collective diffusion coefficients for the lattice shown in Fig. \ref{GaAs4x4}. Unlike earlier approaches, the result will depend on the coverage of the system as our method takes into account inter-particle correlations.

As usual, we start by writing out the variational formula for the lattice studied. As shown in Fig. \ref{GaAs4x4} one elementary cell contains 1 A site, 2 B sites and 2 C sites. There are two types of jumps: $W_{AB}$ and $W_{BC}$. Per elementary cell there are 2 $W_{AB}$ jumps and 4 $W_{BC}$ jumps. Because of that we will have two correlational phases, each assigned to one pair of neighboring sites: $\vec{\Delta}_{AB}$ and $\vec{\Delta}_{BC}$. Depending on the orientation of a given pair in the lattice they will assume the values $\vec{\Delta}_{AB}^{\pm x}=(\pm\Delta_{AB},0)$ and $\vec{\Delta}_{BC}^{\pm x\pm y}=(\pm\Delta_{BC}^{x},\pm\Delta_{BC}^{y})$. Since sites A and C are inversion centers of the lattice, the geometrical phases at those sites are equal to zero. On the other hand, site B does not have such symmetry, however, it has symmetry with respect to the x axis and two inequivalent B sites in an elementary cell are related to each other by the symmetry with respect to the y axis. Because of that the y component or its geometrical phases will vanish, while the x component will differ by the sign between the B sites neighboring to the left and to the right of the A site, that is $\vec{\delta}_{B}^{\pm}=(\pm\delta_{B},0)$. Taking all those facts into account we can write the numerator as
\begin{eqnarray}
M&=&2W_{AB}\theta_{A}(1-\theta_{B})\Big\{\left[(a+\delta_{B}+\theta_{B}\Delta_{AB}+2\theta_{C}\Delta_{BC}^{x})^{2}\right.\nonumber\\
&+&\left.\theta_{B}(1-\theta_{B})\Delta_{AB}^{2}+2\theta_{C}(1-\theta_{C})(\Delta_{BC}^{x})^{2}\right]k_x^2\nonumber\\
&+&2\theta_{C}(1-\theta_{C})(\Delta_{BC}^{y})^{2}k_y^2\Big\}\nonumber\\
&+&4W_{BC}\theta_{B}(1-\theta_{C})\Big\{\left[(\delta_{B}+\theta_{A}\Delta_{AB}+(\theta_{B}+\theta_{C})\Delta_{BC}^{x})^{2}\right.\nonumber\\
&+&\left.\theta_{A}(1-\theta_{A})\Delta_{AB}^{2}+(\theta_{B}(1-\theta_{B})+\theta_{C}(1-\theta_{C}))(\Delta_{BC}^{x})^{2}\right]k_x^2\nonumber\\
&+&\left[(a-\Delta_{BC}^{y}(\theta_B-\theta_C))^{2}+2(\theta_{B}(1-\theta_{B})\right.\\
&+&\left.\theta_{C}(1-\theta_{C}))(\Delta_{BC}^{y})^{2}\right]k_y^2\Big\}.\nonumber\label{M_GaAs4x4}
\end{eqnarray}
We can see that the diffusion is anisotropic here, however, its main directions are always along x and y axes. The x-component depends on three parameters: $\delta_{B}$, $\Delta_{AB}$ and $\Delta_{BC}^{x}$, while the y-component depends only on one parameter, $\Delta_{BC}^{y}$. One has to minimize both components with respect to those parameters.

For the x-direction we get a set of three equations, which can be again written in the matrix form similar to (\ref{set_3eq})
\begin{equation}
\begin{bmatrix}
X_{11}&X_{12}&X_{13}\\
X_{21}&X_{22}&X_{23}\\
X_{31}&X_{32}&X_{33}\\
\end{bmatrix}
\begin{bmatrix}
\delta_{B}\\
\Delta_{AB}\\
\Delta_{BC}
\end{bmatrix}
=
\begin{bmatrix}
Y_{1}\\
Y_{2}\\
Y_{3}
\end{bmatrix}
\end{equation}
with
\begin{eqnarray}
X_{11}&=&W_{AB}\theta_{A}(1-\theta_{B})+2W_{BC}\theta_{B}(1-\theta_{C})\nonumber\\
X_{22}&=&X_{12}=X_{21}=[W_{AB}(1-\theta_{B})+2W_{BC}(1-\theta_{C})]\theta_{A}\theta_{B}\nonumber\\
X_{33}&=&2[W_{AB}\theta_{A}(1-\theta_{B})\theta_{C}(1+\theta_{C})\nonumber\\
&+&W_{BC}\theta_{B}(1-\theta_{C})(\theta_{B}+\theta_{C}+2\theta_{B}\theta_{C})]\nonumber\\
X_{13}&=&X_{31}=2[W_{AB}\theta_{A}(1-\theta_{B})\theta_{C}\nonumber\\
&+&W_{BC}\theta_{B}(1-\theta_{C})(\theta_{B}+\theta_{C})]\nonumber\\
X_{23}&=&X_{32}=2[W_{AB}(1-\theta_{B})\theta_{C}\nonumber\\
&+&W_{BC}(1-\theta_{C})(\theta_{B}+\theta_{C})]\theta_{A}\theta_{B}\nonumber\\
Y_{1}&=&-aW_{AB}\theta_{A}(1-\theta_{B})\nonumber\\
Y_{2}&=&-aW_{AB}\theta_{A}(1-\theta_{B})\theta_{B}\nonumber\\
Y_{3}&=&-2aW_{AB}\theta_{A}(1-\theta_{B})\theta_{C}
\end{eqnarray}
and again we can obtain the phases
\begin{eqnarray}
\delta_{B}&=&\frac{\det\begin{bmatrix}
	Y_{1}&X_{12}&X_{13}\\
	Y_{2}&X_{22}&X_{23}\\
	Y_{3}&X_{32}&X_{33}\\
	\end{bmatrix}}{\det\begin{bmatrix}
	X_{11}&X_{12}&X_{13}\\
	X_{21}&X_{22}&X_{23}\\
	X_{31}&X_{32}&X_{33}\\
	\end{bmatrix}}\nonumber\\
\Delta_{AB}&=&\frac{\det\begin{bmatrix}
	X_{11}&Y_{1}&X_{13}\\
	X_{21}&Y_{2}&X_{23}\\
	X_{31}&Y_{3}&X_{33}\\
	\end{bmatrix}}{\det\begin{bmatrix}
	X_{11}&X_{12}&X_{13}\\
	X_{21}&X_{22}&X_{23}\\
	X_{31}&X_{32}&X_{33}\\
	\end{bmatrix}}\\
\Delta_{BC}&=&\frac{\det\begin{bmatrix}
	X_{11}&X_{12}&Y_{1}\\
	X_{21}&X_{22}&Y_{2}\\
	X_{31}&X_{32}&Y_{3}\\
	\end{bmatrix}}{\det\begin{bmatrix}
	X_{11}&X_{12}&X_{13}\\
	X_{21}&X_{22}&X_{23}\\
	X_{31}&X_{32}&X_{33}\\
	\end{bmatrix}}.\label{GaAs4x4_phases}\nonumber
\end{eqnarray}
\begin{figure}
	\includegraphics[width=6cm,angle=-90]{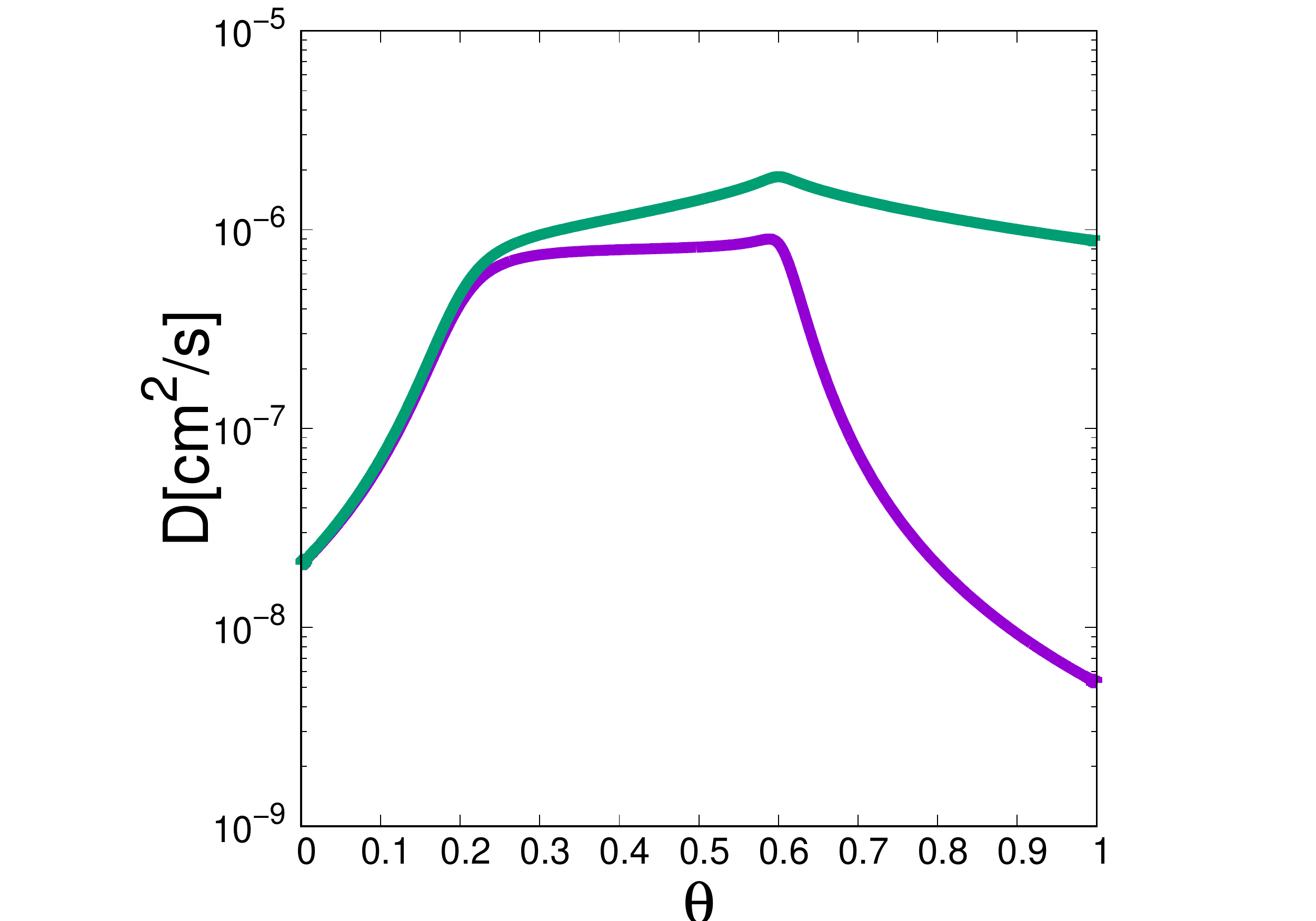}
	\caption{Dependence of the collective diffusion coefficient on the mean coverage of the system for the lattice shown in Fig. \ref{GaAs4x4}. a=0.37 nm, $E_{A}$=0 eV, $E_{B}$=0.18 eV, $E_{C}$=0.48 eV, $E_{AB}$=0.33 eV and $E_{BC}$=0.63 eV. The bottom curve is for the x-direction and the upper one for the y-direction. Temperature T=473K and attempt frequency $\nu=2\cdot10^{13}$/s.}
	\label{GaAs4x4_diff_th}
\end{figure}

The simplified formula (\ref{M_GaAs4x4}) for the x-direction is
\begin{eqnarray}
M_{x}&=&2W_{AB}\theta_{A}(1-\theta_{B})(a+\delta_{B}+\theta_{B}\Delta_{AB}+2\theta_{C}\Delta_{BC}^{x})a,\nonumber\\
\end{eqnarray}
to which we insert the phases (\ref{GaAs4x4_phases}) and obtain the diffusion coefficient
\begin{equation}
D_{x}=\frac{M_{x}}{N},
\end{equation}
where
\begin{equation}
N=\theta_{A}(1-\theta_{A})+2\theta_{B}(1-\theta_{B})+2\theta_{C}(1-\theta_{C}).
\end{equation}

Calculating diffusion coefficient in the y-direction at this lattice is much easier as it requires to minimise the terms in the formula (\ref{M_GaAs4x4}) which stand by $k_y^2$ with respect to only one variational parameter, namely $\Delta_{BC}^y$. After differentiating we get
\begin{eqnarray}
\Delta_{BC}^{y}&=&aW_{BC}\theta_{B}(\theta_{B}-\theta_{C})\Big/\Big\{W_{AB}\theta_{A}(1-\theta_{B})\theta_{C}\\
&+&W_{BC}\theta_{B}[(\theta_{B}-\theta_{C})^{2}+2[\theta_{B}(1-\theta_{B})+\theta_{C}(1-\theta_{C})]]\Big\},\nonumber
\end{eqnarray}
which inserted into simplified formula (\ref{M_GaAs4x4}) for the y-direction gives
\begin{eqnarray}
M_{y}&=&4W_{BC}\theta_{B}(1-\theta_{C})\Big\{1-W_{BC}\theta_{B}(\theta_{B}-\theta_{C})^{2}\Big/\nonumber\\
&&\Big\{W_{AB}\theta_{A}(1-\theta_{B})\theta_{C}+W_{BC}\theta_{B}[(\theta_{B}-\theta_{C})^{2}\nonumber\\
&+&2[\theta_{B}(1-\theta_{B})+\theta_{C}(1-\theta_{C})]]\Big\}\Big\}a^{2}
\end{eqnarray}
and
\begin{equation}
D_{y}=\frac{M_{y}}{N}.
\end{equation}

In Fig. \ref{GaAs4x4_diff_th} we show the collective diffusion coefficients at that lattice as functions of the mean coverage. We take the values of the adsorption energies and the energetic barriers from Ref. \onlinecite{LePage}. We see that both coefficients start at around $2\cdot 10^{-8}$ cm$^{2}$/s for $\theta=0$, which is the value of the same order of magnitude as obtained in the Monte Carlo simulations in Ref. \onlinecite{LePage}. For low coverages the diffusion is almost isotropic. However, starting from $\theta=0.2$ it becomes faster in the y-direction, while in the x-direction it remains constant. At $\theta=0.6$ $D_y$ reaches its maximal value while $D_x$ starts to decrease rapidly. Diffusion in the y-direction becomes two orders of magnitude faster than in the x-direction for the single hole limit ($\theta=1$). Such anisotropy should be easy to verify in STM measurements or other experimental techniques.
\subsubsection{Ga adatoms on GaAs(001)-(2x4) $\beta 2$}
\begin{figure}
	\includegraphics[width=8cm]{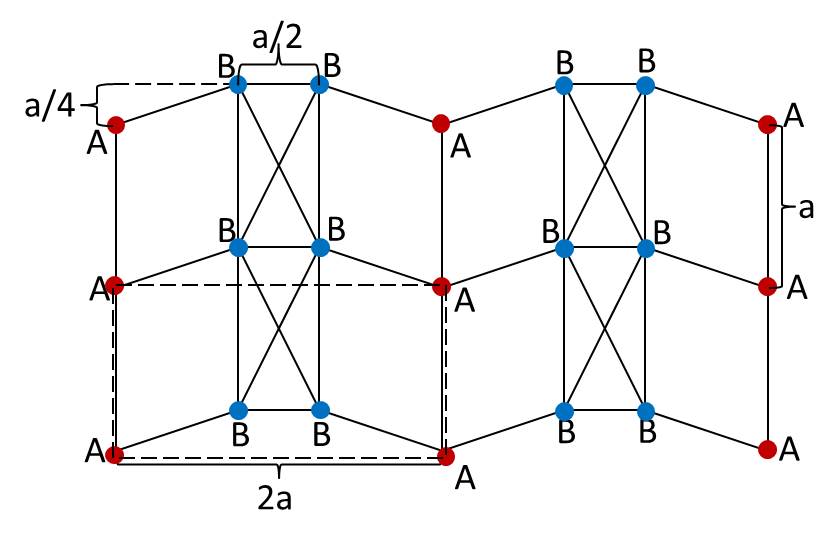}
	\caption{Lattice for Ga adatom diffusing at GaAs(001)-(2x4) $\beta$2, based on Ref. \onlinecite{Kley}. Dashed line shows an elementary cell.}
	\label{GaAs2x4}
\end{figure}

We consider a different reconstruction of the GaAs(001) lattice, namely (2x4) $\beta 2$. We show the scheme of the lattice in Fig. \ref{GaAs2x4}, which was based on ab-initio calculations in Ref. \onlinecite{Kley}.

The lattice has two types of adsorption sites, to which we assign geometrical phases $\vec{\delta}_{A}=(0,\delta_{A})$ and $\vec{\delta}_{B}^{\pm}=(\pm\delta_{B}^{x},\delta_{B}^{y})$, where the sign $\pm$ refers to the B site lying closer either to the left or to the right of the A site. Each cell has one A site and two B sites. There is also one correlational phase $\vec{\Delta}_{AB}^{\pm}=(\pm\Delta_{AB}^{x},\Delta_{AB}^{y})$ assigned to the AB bond, where $\pm$ again refers to the position of the B site to the left or to the right with respect to the A site. The jumps per elementary cell are: one $W_{AA}$, two $W_{AB}$, one $W_{BB}^{h}$ (horizontal), two $W_{BB}^{v}$ (vertical) and two $W_{BB}^{d}$ (diagonal). We write the numerator as
\begin{eqnarray}
M&=&W_{AA}\theta_{A}(1-\theta_{A})\{4\theta_{B}(1-\theta_{B})(\Delta_{AB}^{x})^{2}k_x^2\label{M_GaAs2x4}\\
&+&[a^2+4\theta_{B}(1-\theta_{B})(\Delta_{AB}^{y})^{2}]k_y^2\}\nonumber\\
&+&2W_{AB}\theta_{A}(1-\theta_{B})\Bigg\{\bigg[\left(\frac{3}{4}a+\delta_{B}^{x}+\theta_{B}\Delta_{AB}^{x}\right)^{2}\nonumber\\
&+&\theta_{B}(1-\theta_{B})(\Delta_{AB}^{x})^{2}\bigg]k_x^2+\bigg[\left(\frac{1}{4}a+\delta_{B}^{y}-\delta_{A}-\theta_{B}\Delta_{AB}^{y}\right)^2\nonumber\\
&+&\theta_{B}(1-\theta_{B})(\Delta_{AB}^{y})^2\bigg]k_y^2\Bigg\}\nonumber\\
&+&(W_{BB}^{h}+2W_{BB}^{d})\theta_{B}(1-\theta_{B})\Bigg\{\bigg[\left(\frac{1}{2}a-2\delta_{B}^{x}-2\theta_{A}\Delta_{AB}^{x}\right)^2\nonumber\\
&+&2\theta_{A}(1-\theta_{A})(\Delta_{AB}^{x})^{2}\bigg]k_x^2+2\theta_{A}(1-\theta_{A})(\Delta_{AB}^{y})^{2}k_y^2\Bigg\}\nonumber\\
&+&2(W_{BB}^{d}+W_{BB}^{v})\theta_{B}(1-\theta_{B})a^2k_y^2\nonumber\\
&+&4W_{BB}^v\theta_{A}(1-\theta_{A})\theta_{B}(1-\theta_{B})[(\Delta_{AB}^{x})^{2}k_x^2+(\Delta_{AB}^{y})^2k_y^2].\nonumber
\end{eqnarray}
Again one has to minimize the above expression separately for the x- and the y-direction with respect to the variational parameters. For the x-direction we have
\begin{eqnarray}
\delta_{B}^{x}&=&a\Big\{(W_{AA}+W_{BB}^{v})(1-\theta_{A})\Big[(W_{BB}^{h}+2W_{BB}^{d})\theta_{B}\nonumber\\
&-&\frac{3}{2}W_{AB}\theta_{A}\Big]+\frac{1}{2}\Big[W_{AB}+W_{BB}^{h}+2W_{BB}^{d}\Big]\label{phases_GaAs2x4}\\
&\times&\Big[(W_{BB}^{h}+2W_{BB}^{d})\theta_{B}(1-\theta_{A})-\frac{3}{2}W_{AB}\theta_{A}(1-\theta_{B})\Big]\nonumber\\
&+&\frac{3}{4}W_{AB}(W_{BB}^{h}+2W_{BB}^{d})\theta_{A}(\theta_{B}-\theta_{A})\Big\}\Big/\nonumber\\
&&\Big\{(W_{AB}+W_{BB}^{h}+2W_{BB}^{d})[W_{AB}\theta_{A}(1-\theta_{B})\nonumber\\
&+&2(W_{BB}^{h}+2W_{BB}^{d})\theta_{B}(1-\theta_{A})]\nonumber\\
&+&W_{AB}(W_{BB}^{h}+2W_{BB}^{d})\theta_{A}(\theta_{A}-\theta_{B})\nonumber\\
&+&2(W_{AA}+W_{BB}^{v})(1-\theta_{A})[W_{AB}\theta_{A}\nonumber\\
&+&2(W_{BB}^{h}+2W_{BB}^{d})\theta_{B}]\Big\},\nonumber\\
\Delta_{AB}^{x}&=&2a(\theta_{A}-\theta_{B})W_{AB}(W_{BB}^{h}+2W_{BB}^{d})\Big/\nonumber\\
&&\Big\{(W_{AB}+W_{BB}^{h}+2W_{BB}^{d})[W_{AB}\theta_{A}(1-\theta_{B})\nonumber\\
&+&2(W_{BB}^{h}+2W_{BB}^{d})\theta_{B}(1-\theta_{A})]\nonumber\\
&+&W_{AB}(W_{BB}^{h}+2W_{BB}^{d})\theta_{A}(\theta_{A}-\theta_{B})\nonumber\\
&+&2(W_{AA}+W_{BB}^{v})(1-\theta_{A})[W_{AB}\theta_{A}\nonumber\\
&+&2(W_{BB}^{h}+2W_{BB}^{d})\theta_{B}]\Big\}.\nonumber
\end{eqnarray}
The corresponding diffusion coefficient is
\begin{eqnarray}
D_{x}&=&a\Big\{\frac{3}{2}W_{AB}\theta_{A}(1-\theta_{B})\left(\frac{3}{4}a+\delta_{B}^{x}+\theta_{B}\Delta_{AB}^{x}\right)\nonumber\\
&+&\frac{1}{2}(W_{BB}^{h}+2W_{BB}^{d})\theta_{B}(1-\theta_{B})\Big(\frac{1}{2}a-2\delta_{B}^{x}\\
&-&2\theta_{A}\Delta_{AB}^{x}\Big)\Big\}\Big/\Big\{\theta_{A}(1-\theta_{A})+2\theta_{B}(1-\theta_{B})\Big\}.\nonumber
\end{eqnarray}
\begin{figure}
	\includegraphics[width=6cm,angle=-90]{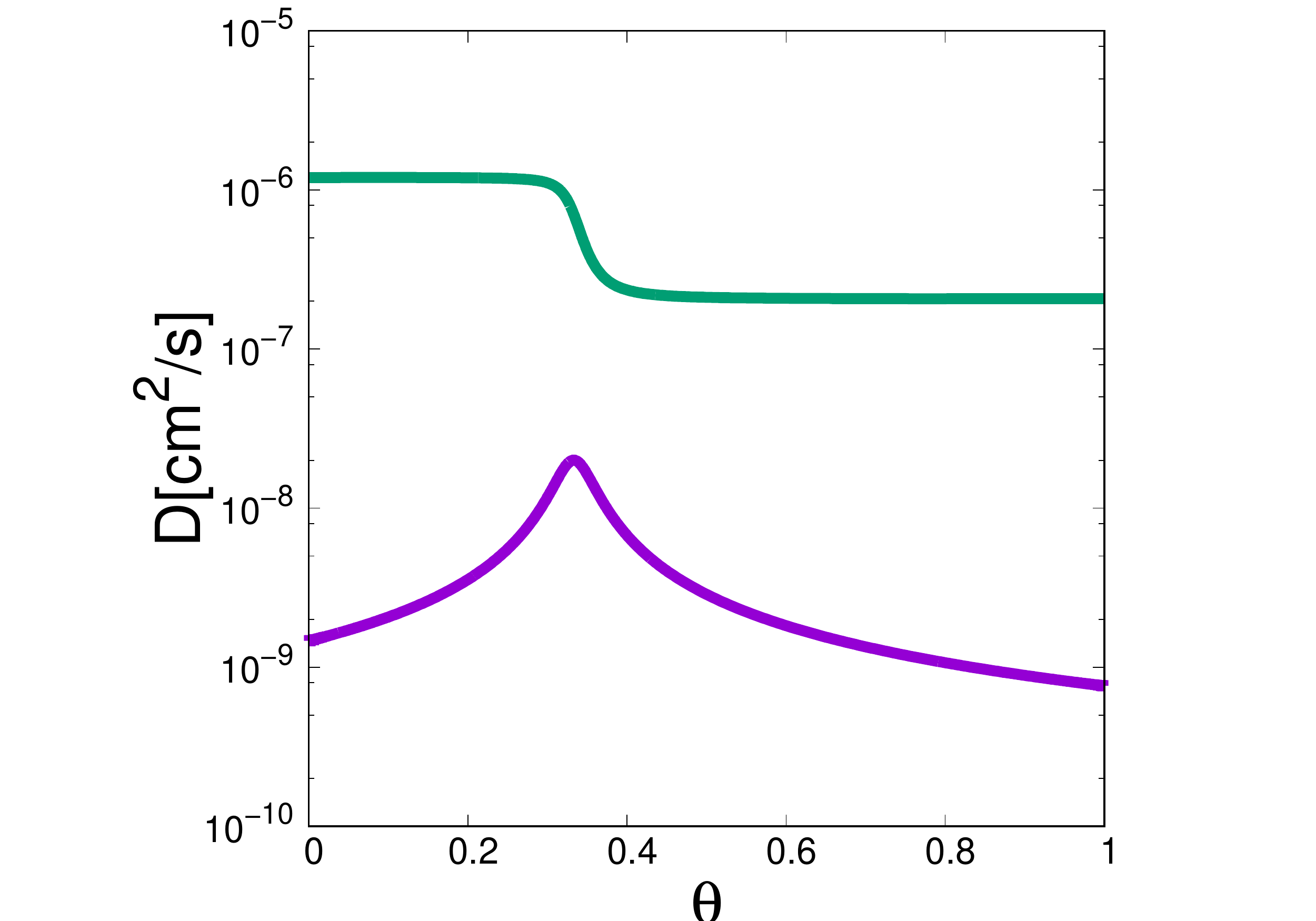}
	\caption{Dependence of the collective diffusion coefficient on the mean coverage of the system for the lattice shown in Fig. \ref{GaAs2x4}. a=1.13 nm, $E_{A}$=0 eV, $E_{B}$=0.3 eV, $E_{AB}$=0.8 eV, $E_{BB}^{h}$=0.7 eV and $E_{BB}^{v}=E_{BB}^{d}$=0.9 eV. The bottom curve is for the x-direction and the upper one for the y-direction. Temperature T=473K and attempt frequency $\nu=2\cdot10^{13}$/s.}
	\label{GaAs2x4_diff_th}
\end{figure}

In the above we used the denominator $N=\theta_{A}(1-\theta_{A})+2\theta_{B}(1-\theta_{B})$ and the variational parameters should be taken from Eq. (\ref{phases_GaAs2x4}). The obtained formula is valid for any value of the coverage $\theta$. For $\theta\rightarrow 0$ we have $\Delta_{AB}^{x}=0$ but $\delta_{B}^{x}\neq 0$ and
\begin{equation}
D_{x}=\frac{4W_{AB}W_{BA}(W_{BB}^{h}+2W_{BB}^{d})}{(2W_{AB}+W_{BA})(W_{BA}+2W_{BB}^{h}+4W_{BB}^{d})}a^2,
\end{equation}
which is the same expression as the one obtained in Ref. \onlinecite{Gosalvez2}.

Now let us consider diffusion in the y-direction. There are three variational phases related to that direction: $\delta_{A}$, $\delta_{B}^{y}$ and $\Delta_{AB}^{y}$. However, they enter the numerator (\ref{M_GaAs2x4}) only as the difference $\delta_{B}^{y}-\delta_{A}$. Therefore, we can use that difference as a new parameter $\delta_{AB}^{y}$ reducing the number of independent variational phases to two. After differentiating we get
\begin{eqnarray}
\delta_{AB}^{y}&=&-\frac{a}{4}\nonumber\\
\Delta_{AB}^{y}&=&0.
\end{eqnarray}
Even though we allowed for a non-zero value of $\Delta_{AB}^{y}$ it turned out to be zero. It reflects the fact that the two channels of diffusion in the y-direction, one through A sites and the other through B sites, are independent from each other. The final result for the diffusion coefficient is
\begin{eqnarray}
D_{y}&=&a^{2}\frac{W_{AA}\theta_{A}(1-\theta_{A})+2(W_{BB}^{v}+W_{BB}^{d})\theta_{B}(1-\theta_{B})}{\theta_{A}(1-\theta_{A})+2\theta_{B}(1-\theta_{B})}.\nonumber\\
\end{eqnarray}
Since the contributions to the diffusion in the y-direction from the $W_{AB}$ jumps cancel out, the above expression does not depend on that jump rate. It is easy to see that in the $\theta\rightarrow 0$ limit one again gets the result obtained in Ref. \onlinecite{Gosalvez2}.

Again we illustrate the results by plotting the dependence of the diffusion coefficients as functions of the coverage in Fig. \ref{GaAs2x4_diff_th} with the values of the barriers taken from Ref. \onlinecite{Kley}. We can see that there is a high anisotropy in the whole range of the coverage. The behaviors of diffusion in the two directions are completely distinct. Diffusion along x-direction is mostly through three adsorption sites: one deeper (A) and two shallow ones (B). Therefore, it is similar to the one-dimensional case discussed in Ref. \onlinecite{Minkowski5}. There is a characteristic peak at $\theta=1/3$ related to ordering of the system at that coverage, in which sites A are completely filled while sites B are free. On the other hand, diffusion along y-direction exhibits a behavior of switching between the deep and the shallow diffusive channel. Such behavior was earlier observed in the model of hills and valleys \cite{Minkowski3}, where the number of sites in the deep channel and the shallow one was equal and the transition between them occurred at $\theta=1/2$ instead of $\theta=1/3$.

\section{Conclusions}
We have formulated in a simple way a variational method of analyzing collective diffusion of particles adsorbed at lattices of arbitrary geometry, interacting by infinite on-site repulsion, excluding double site occupancy. We have derived a convenient variational formula which is expressed by all jump rates and equilibrium occupancies of the adsorption sites in a given system, that is its energetic landscape. Finding diffusion coefficients from that formula is a simple and straightforward procedure. It has been shown how the method can be applied to various systems, including real ones with specific values of the energetic barriers.

We showed explicit calculations of the diffusion coefficient based on our variational approach for a few lattices taken from Refs \onlinecite{Gosalvez1} and \onlinecite{Gosalvez2}. In each case our results reproduce the expressions for the single-particle diffusion coefficient (taking the limit $\theta\rightarrow 0$) calculated in those publications. Our approach allows to find the expressions for the diffusion coefficient for any value of the system's coverage $\theta$. For the values of $\theta$ between 0 and 1 the effects of inter-particle correlations become important and the collective diffusion coefficient is different from the one obtained without assuming correlations in the system. Only in the case with identical adsorption energies of all sites the diffusion coefficient does not depend on the coverage \cite{Kutner,Minkowski1}. However, in real systems that condition is rarely fulfilled.

The presented method was applied to two different reconstructions of GaAs(001) surface and we can study the coverage dependence of the diffusion coefficient. Increasing Ga coverage not only changes value of the diffusion coefficient, but also dramatically influences the ratio of diffusion in two main directions. At the surface of c(4x4) reconstruction diffusion changes from isotropic for low Ga coverages into strongly one-dimensional for high Ga coverages, for which diffusion coefficients in two directions differ by three orders of magnitude. At the surface of (2x4) $\beta$2 reconstruction symmetry diffusion of Ga adatoms is faster in one direction for all coverages, but ratio between $D_x$ and $D_y$ changes from three orders of magnitude, then decreases to one for coverage 0.35 and for higher coverages increases again. Diffusion is never isotropic for this reconstruction of GaAs(001) surface.

It is important to note that the approach presented in this article is not limited to the examples shown in the section with the results. The variational method is not dependent on the symmetry of the lattice, it is the variational parameters whose values adjust to the geometry of the energetic landscape and the correlations between the diffusing particles to give a proper expression for the diffusion coefficient. Our approach is also not limited to two dimensions, without any modifications in the formalism one can apply it also to analyze bulk diffusion.
\begin{acknowledgments}
The research was supported by the National Science Centre (NCN) of Poland (NCN Grant No. 2015/17/N/ST3/02310).
\end{acknowledgments}

\end{document}